\documentclass[oldversion]{aa}  
\usepackage{graphicx,amsmath}
\usepackage{amssymb}
\usepackage{color}
\usepackage{natbib}		
\usepackage{url}
\usepackage{multirow}
\usepackage{threeparttable}   
\usepackage{soul}
\usepackage{xcolor}
 
\bibpunct{(}{)}{;}{a}{}{,} 
\defcitealias{Zhang2017}{ZS}
\usepackage[version=3]{mhchem}

\begin{document}

   \title{Hot Exoplanet Atmospheres Resolved with Transit Spectroscopy (HEARTS)\thanks{Based on observations made at ESO 3.6 m telescope (La Silla, Chile) under ESO programme 100.C-0750. The custom CCF mask built for WASP-121, as well as the EulerCam lightcurves, are available at the CDS via anonymous ftp to cdsarc.u-strasbg.fr (130.79.128.5) or via http://cdsarc.u-strasbg.fr/viz-bin/qcat?J/A+A/} \\
   III. Atmospheric structure of the misaligned ultra-hot Jupiter WASP-121b}

   \author{
   V. Bourrier\inst{1},
   D. Ehrenreich\inst{1},
   M. Lendl\inst{1},  
   M. Cretignier\inst{1},
   R. Allart\inst{1},
   X. Dumusque\inst{1},     
   H.M Cegla\inst{1},       
   A. Su\'arez-Mascare\~no\inst{2},
   A. Wyttenbach\inst{3},     
   H.J. Hoeijmakers\inst{1,4},         
   C. Melo\inst{5},        
   T. Kuntzer\inst{1},    
   N. Astudillo-Defru\inst{6},
   H. Giles\inst{1}, 
   K. Heng\inst{2},
   D. Kitzmann\inst{4}, 
   B. Lavie\inst{1},  
   C. Lovis\inst{1}, 
   F. Murgas\inst{2,7}, 
   V. Nascimbeni\inst{8},
   F. Pepe\inst{1}, 
   L. Pino\inst{9},
   D. Segransan\inst{1}, 
   S. Udry\inst{1}   
	}

\authorrunning{V.~Bourrier et al.}
\titlerunning{Atmospheric and orbital characterization of WASP-121b}

\offprints{V.B. (\email{vincent.bourrier@unige.ch})}

\institute{
Observatoire de l'Universit\'e de Gen\`eve, 51 chemin des Maillettes, 1290 Versoix, Switzerland     
\and 		
Instituto de Astrofísica de Canarias (IAC), E-38205 La Laguna, Tenerife, Spain
\and
Leiden Observatory, Leiden University, Postbus 9513, 2300 RA Leiden, The Netherlands
\and
Center for Space and Habitability, Universit\"at Bern, Gesellschaftsstrasse 6, 3012 Bern, Switzerland     
\and
European Southern Observatory, Alonso de C\'ordova 3107, Vitacura, Región Metropolitana, Chile
\and        
Departamento de Matem\'atica y F\'isica Aplicadas, Universidad Cat\'olica de la Sant\'isima Concepci\'on, Alonso de Rivera 2850, Concepci\'on, Chile 
\and       
Departamento de Astrof\'sica, Universidad de La Laguna (ULL), E-38206 La Laguna, Tenerife, Spain
\and
INAF – Osservatorio Astronomico di Padova, Vicolo dell’Osservatorio 5, 35122, Padova, Italy
\and
Anton Pannekoek Institute for Astronomy, University of Amsterdam, Science Park 904, 1098 XH Amsterdam, The Netherlands
} 
   
   \date{} 
 
  \abstract
{Ultra-hot Jupiters offer interesting prospects for expanding our theories on dynamical evolution and the properties of extremely irradiated atmospheres. In this context, we present the analysis of new optical spectroscopy for the transiting ultra-hot Jupiter WASP-121b. We first refine the orbital properties of WASP-121b, which is on a nearly polar (obliquity $\psi^{\rm North}$=88.1$\pm$0.25$^{\circ}$ or $\psi^{\rm South}$=91.11$\pm$0.20$^{\circ}$) orbit, and exclude a high differential rotation for its fast-rotating (P$<$1.13\,days), highly inclined ($i_\mathrm{\star}^{\rm North}$=8.1$\stackrel{+3.0}{_{-2.6}}^{\circ}$ or $i_\mathrm{\star}^{\rm South}$=171.9$\stackrel{+2.5}{_{-3.4}}^{\circ}$) star. We then present a new method that exploits the reloaded Rossiter-McLaughlin technique to separate the contribution of the planetary atmosphere and of the spectrum of the stellar surface along the transit chord. Its application to HARPS transit spectroscopy of WASP-121b reveals the absorption signature from metals, likely atomic iron, in the planet atmospheric limb. The width of the signal (14.3$\pm$1.2\,km\,s$^{-1}$) can be explained by the rotation of the tidally locked planet. Its blueshift (-5.2$\pm$0.5\,km\,s$^{-1}$) could trace strong winds from the dayside to the nightside, or the anisotropic expansion of the planetary thermosphere.}


   \maketitle


\section{Introduction}

Ultra-hot Jupiters allow us to advance our understanding of planetary migration and orbital stability (\citealt{Delrez2016}), and they offer great prospects for atmospheric characterization (\citealt{Parmentier2018}). Their high temperature (typically higher than about 2000\,K) simplifies the atmospheric chemistry by dissociating molecular species into their atomic constituents (\citealt{Lothringer2018}). Multiple atoms and ions could thus be detected in the atmospheric limb of ultra-hot Jupiters using high-resolution transmission spectroscopy (e.g., \citealt{Hoeijmakers2018, Hoeijmakers2019} for the prototypical ultra-hot Jupiter KELT-9b). These planets are also interesting candidates for probing evaporation and the effect of photo-ionization on the upper atmospheric structure. Their extreme irradiation by the host star causes the hydrodynamical expansion of their upper atmosphere, allowing metals to escape and be detected in the near-ultraviolet after they are ionized in the exosphere (\citealt{Fossati2010}, \citealt{Haswell2012}, \citealt{Sing2019}).  

Interestingly, several ultra-hot Jupiters were found on highly misaligned orbits (e.g. WASP-12b, \citealt{Albrecht2012}; WASP-33b, \citealt{Cameron2010}; WASP-121b, \citealt{Delrez2016}), suggesting dynamical migration processes induced by gravitational interactions with companions, rather than disk migration (e.g. \citealt{Nagasawa2008}, \citealt{Fabrycky2007}, \citealt{Guillochon2011}). At such close distances to their stars, ultra-hot Jupiters are subjected to strong tidal interactions that determine their final orbital evolution. Precisely measuring the orbital architecture of ultra-hot Jupiters and monitoring its evolution is thus of particular importance to determine their migration history and their potential decay into the star. The occultation of a rotating star by a transiting planet removes the light of the hidden photosphere from the observed stellar lines (the so-called Rossiter-McLaughlin effect, or RM effect, \citealt{Holt1893}; \citealt{Rossiter1924}; \citealt{McLaughlin1924}). Different techniques have been developed to analyze the radial velocity (RV) anomaly induced by the distortion of the stellar absorption lines (e.g., \citealt{ohta2005}, \citealt{gimenez2006}, \citealt{hirano2011b}, \citealt{boue2013}), to model their profile while accounting for the planet occultation (e.g., \citealt{Cameron2010}, \citealt{Gandolfi2012}, \citealt{Crouzet2017}), or to isolate the local stellar lines from the planet-occulted regions (e.g., \citealt{Cegla2016}, \citealt{Bourrier_2018_Nat}). These techniques enable deducing the trajectory of the planet across the stellar disk, and thus inferring the projected or true 3D alignment between the spins of the planetary orbit and the stellar rotation. \\

The ultra-hot Jupiter WASP-121b (\citealt{Delrez2016}) is a good candidate for both atmospheric and orbital architecture studies (Table~\ref{tab:sys_prop}). This super-inflated gas giant transits a bright F6-type star (V = 10.4), favoring optical transmission spectroscopy measurements. Its near-polar orbit at the edge of the Roche limit ($P$ = 1.27\,days) makes WASP-121b subject to strong tidal interactions with the star (\citealt{Delrez2016}) and an intense atmospheric escape. The increase in transit depth of WASP-121b toward near-UV wavelengths (\citealt{Evans2018}, \citealt{Salz2019_NUV_WASP121b}) was recently shown to arise from iron and magnesium atoms that escape into the exosphere (\citealt{Sing2019}), which confirms the hydrodynamical evaporation of WASP-121b and opens new avenues to link the structure and composition of the lower and upper atmosphere.

In the present study we investigate the atmosphere of WASP-121b, and refine the properties of its planetary system. In Sect.~\ref{sec:RV_fit}, we reanalyze long-term RV and activity indexes of the system. Sect.~\ref{sec:reloaded RM} exploits transit spectroscopy of WASP-121b obtained with the High Accuracy Radial velocity Planet Searcher (HARPS), combined with simultaneous EulerCam photometry, to analyze the orbital architecture of WASP-121b and its star. In Sect.~\ref{sec:atmo_struc} we characterize the atmospheric structure of the planet at the limb, using a new method to isolate the signal of the planetary atmosphere from the occulted stellar lines. We conclude the study in Sect.~\ref{sec:conclu}.



\section{Radial velocity monitoring of WASP-121}
\label{sec:RV_fit}

\subsection{Planet-induced motion}

\begin{center}
\begin{figure}
\centering
\includegraphics[trim=0.cm 0.cm 0.cm 0cm,clip=true,width=\columnwidth]{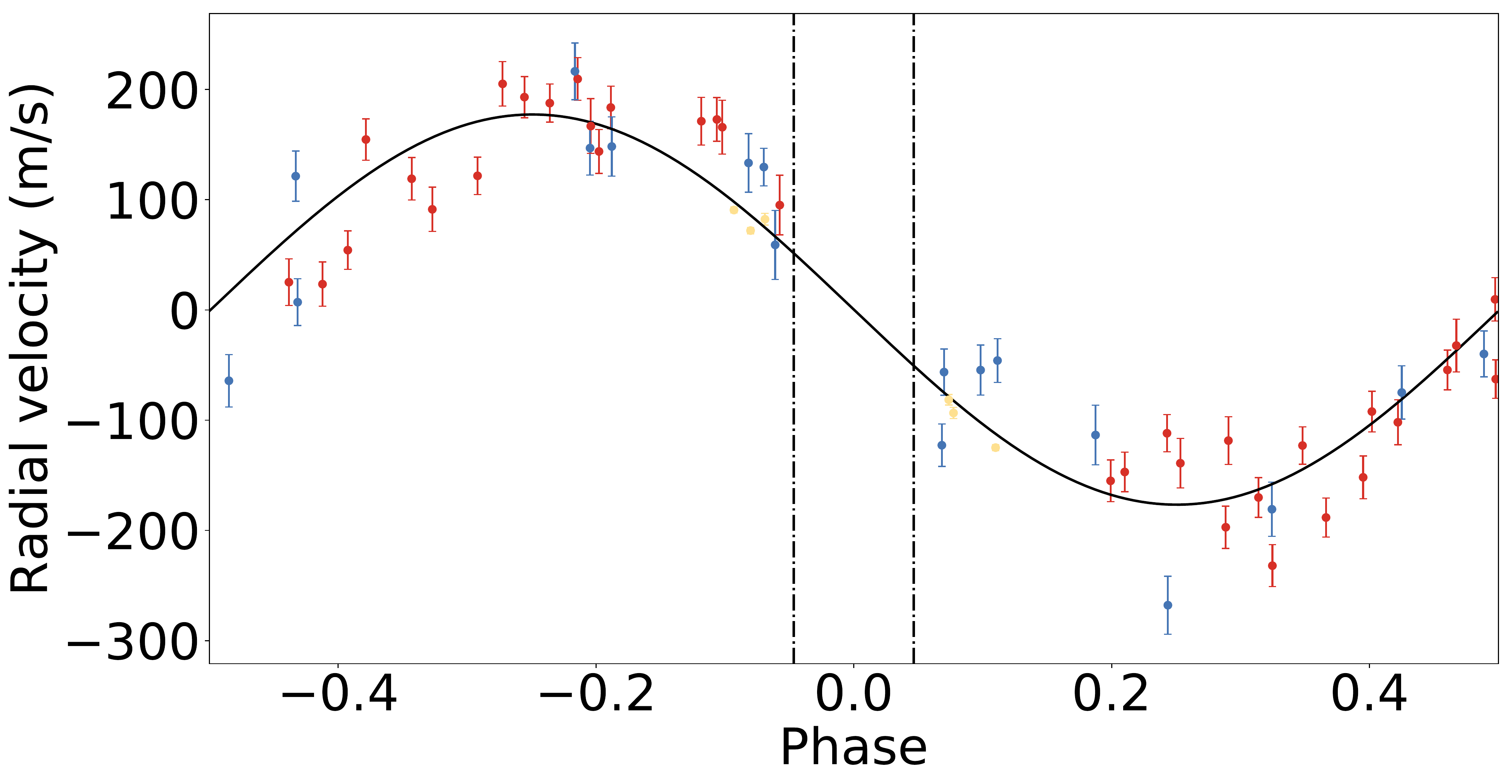}
\caption[]{Radial velocities of WASP-121b, phase-folded with the orbital period and detrended from stellar activity. Coralie data points have been obtained before (red) and after (blue) the fiber upgrade (\citealt{Segransan2010}). HARPS datapoints are shown in gold. Our best-fit Keplerian model to the out-of-transit data is shown as a solid black line (datapoints obtained in the frame of transit observations were binned, see text). The first and fourth transit contacts are indicated by dash-dotted vertical black lines. }
\label{fig:RV_fit}
\end{figure}
\end{center}

We analyzed RV data points of WASP-121 obtained with the Coralie (\citealt{baranne1996}, \citealt{Queloz2000}) and HARPS (\citealt{Mayor2003}) spectrographs to revise the semi-amplitude of the stellar reflex motion and the mass of WASP-121b (the complete RV dataset is shown in Fig.~\ref{fig:RV_ana_appendix_nobin}). RV data were analyzed with the Data and Analysis Center for Exoplanets web platform (DACE\footnote{https://dace.unige.ch}). We excluded datapoints obtained during four planet transits (one observed with Coralie, the other three with HARPS) and binned the remaining data in time by 0.25 day separately for each instrument (to mitigate short-term stellar signals and to avoid favorring HARPS datapoints). The processed data (Fig.~\ref{fig:RV_fit}) were fit with the Keplerian model described in \citet{Delisle2016}. It was combined with the activity detrending described in \citet{Delisle2018}, which adds a term that is linearly correlated with the bisector of the cross-correlation functions (CCFs). The model was fit to the data using a Markov chain Monte Carlo (MCMC) algorithm (\citealt{Diaz2014,Diaz2016}) with Gaussian priors on the period, time of mid-transit, eccentricity, and periastron argument derived from photometry obtained with the Transiting Exoplanet Survey Satellite (TESS) by \citet{Bourrier2019}. Results are given in Table~\ref{fig:RV_ana_appendix}. The mass we derive for WASP-121b is consistent with that of \citet{Delrez2016}. We kept the values of the properties that have been derived from the TESS photometry as our final estimates for the revised planetary properties (Table~\ref{tab:sys_prop}), because the fit to the RV data did not improve their precision, nor changed their values significantly.\\

\subsection{Stellar rotation}
\label{sec:Prot}

\begin{figure*}[]
\center
\includegraphics[angle=0,width=16cm]{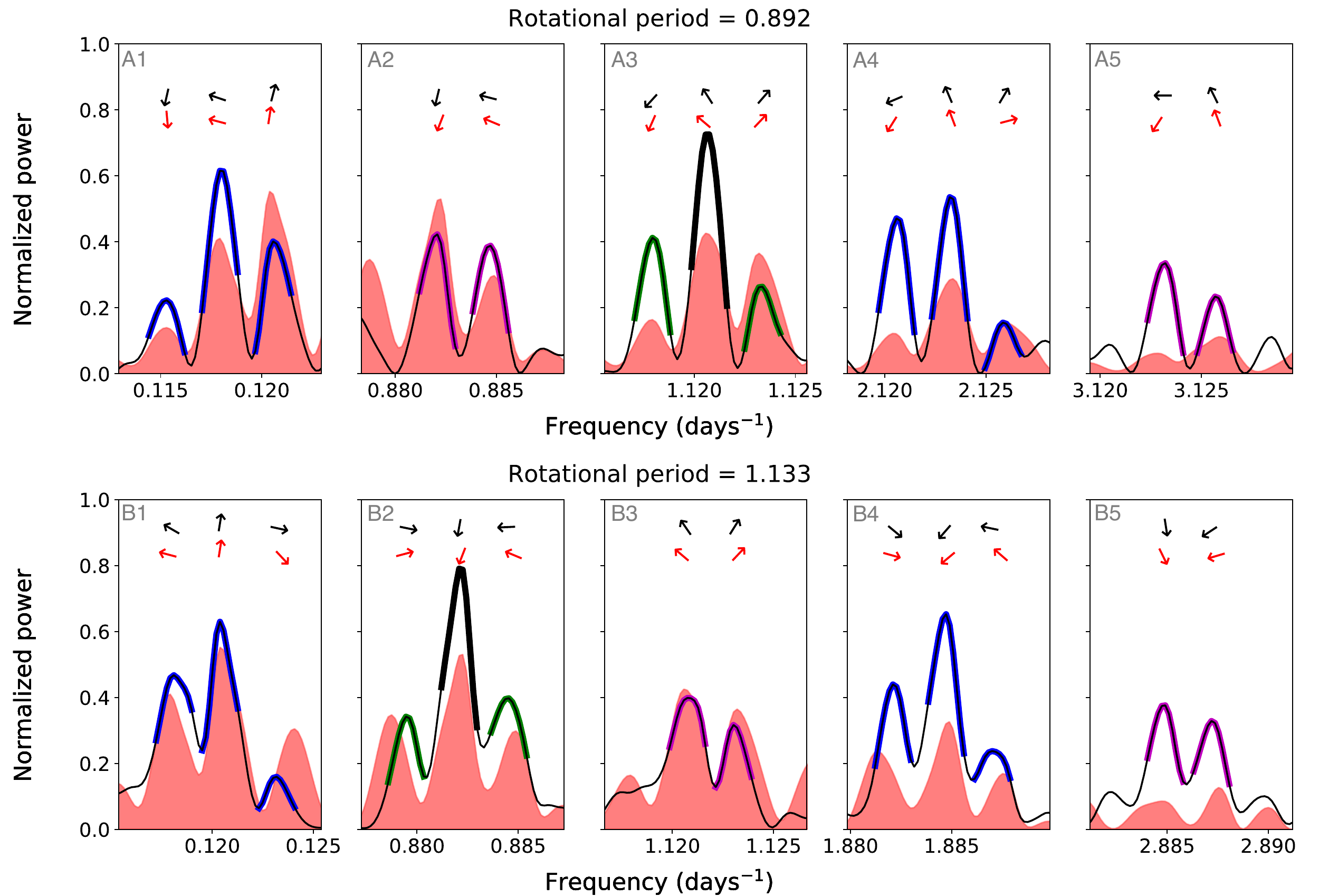}
\caption[]{Comparison between the real and alias signals in the RV data of WASP-121 corrected for the planet signal (red shadow) and in the simulated data (black). The top panel corresponds to the 0.89-day signal, the bottom panel to the 1.13-day signal. These plots correspond to one realization of noise out of 100 different trials. Real signals are shown in black (panels A3 and B2), yearly aliases ($\pm$0.0027 days$^{-1}$) in green (panels A3 and B2), daily aliases ($\pm$1, $\pm$1.0027 and $\pm$1.0056 days$^{-1}$) in blue (panels A1, A4, B1, and B4), and 2-day aliases ($\pm$2.0027 and $\pm$2.0055 days$^{-1}$) in purple (panels A2, A5, B3, and B5). Arrows at the top of each peak show the phase of each signal for the real (red) and the simulated data (black).}
\label{fig:1}
\end{figure*}

After the contribution of WASP-121b was removed, the periodogram of the RV residuals reveals three significant signals at periods of 0.89, 1.13, and $\sim$8.4 days. These signals are also visible when periodograms of the bisector span (BIS SPAN) and the full-width at half-maximum (FWHM) time-series measured on the CCF are analyzed. They arise from magnetically active regions at the surface of fast-rotating WASP-121, and are all aliases of one another. We show in Sect.~\ref{sec:results_RM} that the signal at $\sim$8.4 days must be an alias because of the high measured stellar projected rotational velocity. We then used the technique proposed by \citet{Dawson2010} to determine whether the signals at 0.89 or 1.13 days directly trace the rotational modulation of WASP-121. To distinguish the real signal and aliases, \citet{Dawson2010} proposed simulating data with the same time sampling and injecting the signals that are to be tested as being real or aliases. For each injected signal, a comparison of the period and phase of all the aliases created by the observational sampling is then performed between the simulated and real data set. Using this technique, \citet{Dawson2010} were able to show that the period originally derived for planet 55\,Cnc\,e (\citealt{McArthur2004}, \citealt{Fischer2008}) was an alias of the real signal.\\
Here, we extend the approach of \citet{Dawson2010} by performing 100 simulations for each injected signal, taking different configurations for the noise into account. We also analyze the rotational signal using the RVs, the BIS SPAN and the FWHM time-series. For each real or alias signal in the real or simulated data, we calculate the area below each peak and its phase (Fig.\,\ref{fig:1}). The area is defined as the sum of the power for all frequencies that lie 5 bins away from the frequency corresponding to the maximum power of the peak. Finally, the sum of both the absolute phase and area differences are calculated for each of the 100 simulations on the RVs, the BIS SPAN and the FWHM. These sums are given in Table\,\ref{table:rotation_period}. Overall, we observe smaller differences between the real and simulated data when the 1.13-day signal is considered compared to the signal at 0.89 day. We therefore propose that the 1.13-day signal traces the rotational modulation of WASP-121. \\

\begin{table}[tbh]
\caption{
\label{table:rotation_period}
Area and phase differences, in arbitrary units, for the 0.89- and 1.13-day signals seen in the RV, BIS SPAN, and FWHM time-series periodograms. Bold numbers highlight the lower difference values when the two signals are compared.}
\begin{center}
\begin{tabular}{cccc}
\hline\hline       
    &    Period [d]  					& Area difference    & Phase difference   \\
\hline
RV  &  \begin{tabular}{@{}c@{}}0.89 \\ 1.13\end{tabular}   &   \begin{tabular}{@{}c@{}}922 \\ \bf{696}\end{tabular} &  \begin{tabular}{@{}c@{}}1629 \\ \bf{1347}\end{tabular} \\
\hline
BIS SPAN  &  \begin{tabular}{@{}c@{}}0.89 \\ 1.13\end{tabular}   &   \begin{tabular}{@{}c@{}}1923 \\ \bf{690}\end{tabular} &  \begin{tabular}{@{}c@{}}\bf{447} \\ 647\end{tabular} \\
\hline
FWHM  &  \begin{tabular}{@{}c@{}}0.89 \\ 1.13\end{tabular}   &   \begin{tabular}{@{}c@{}}1975 \\ \bf{1202}\end{tabular} &  \begin{tabular}{@{}c@{}}\bf{2608} \\ 2820\end{tabular} \\
\hline
\end{tabular}
\end{center}
\end{table}

\begin{figure}
\begin{center}
\includegraphics[trim=1.5cm 0.8cm 1.cm 0.5cm,clip=true,width=\columnwidth]{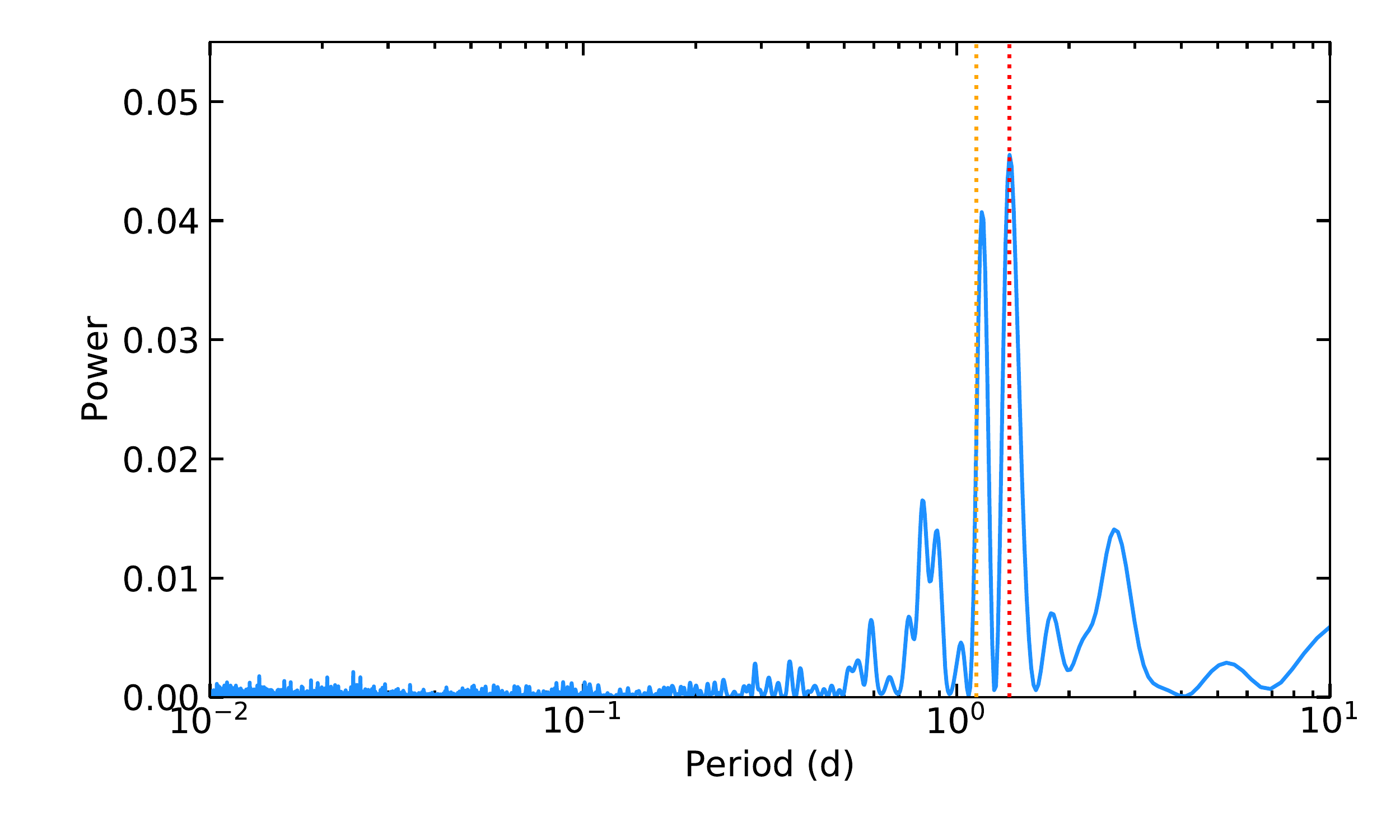}
\caption{\label{fig:TESS:residualspg} Lomb-Scargle periodogram of the residuals between TESS photometry in orbit 22 and the best-fit model for WASP-121b. The orange dashed line indicates the rotational modulation identified in the RVs at 1.13\,d. The peak indicated by the red dashed line at 1.37\,d is only present in this orbit, and likely traces a transient active region at higher stellar latitudes.}
\end{center}
\end{figure}

We also ran a periodogram analysis on the residuals between the TESS photometry and the best-fit model derived by \citet{Bourrier2019}. The two strongest peaks are measured at periods of 1.16 and 1.37 days. The first signal corresponds well to the rotational modulation identified in the RV of WASP-121, and likely originates in the same active regions at the surface of the star. WASP-121 was observed over two TESS orbits. We cut each of them in half, and ran independent periodogram analyses on the four resulting segments. The stronger 1.37-day signal is only present in the second TESS orbit, with similar power in its two halves (Fig.~\ref{fig:TESS:residualspg}). Our best interpretation is that WASP-121 rotates differentially, with the 1.37-day signal arising from active regions located at higher latitudes (and thus rotating slower) than those responsible for the 1.13-day signal. These high-latitude regions would have developed rapidly around epoch $\sim 1502$ and lasted at least for the rest of the TESS observations. The possibility for differential rotation is investigated in more detail in Sect.~\ref{sec:fit_RM}. \\

\begin{table*}   
\caption[]{Properties of the WASP-121 system}
\centering
\begin{threeparttable}
\begin{tabular}{c|c|c|c|c}
    \hline
    \hline
      Parameter & Symbol & Value  &  Unit   &   Origin  \\
    \hline      
\textit{Stellar properties}    &   &   &   &  \\      
    \hline   
Mass		& $M_{\star}$ & 1.358$\stackrel{+0.075}{_{-0.084}}$   & M$_{\odot}$  &  \citealt{Delrez2016} \\
Radius		& $R_{\star}$ & 1.458$\pm$0.030  & R$_{\odot}$  &  \citealt{Delrez2016} \\    
Density		& $\rho_{\star}$ & 0.434$\pm$0.038  &  $\rho_{\odot}$ & \citealt{Delrez2016}$^{\dagger}$ \\
Limb-darkening coefficients & $u_{1}$ & 0.364$\stackrel{+0.034}{_{-0.030}}$   & &  EulerCam \\
			& $u_{2}$ & 0.146$\stackrel{+0.066}{_{-0.049}}$ & &   EulerCam    \\
Inclination   &  $i_\mathrm{\star}^{\rm North}$   &  8.1$\stackrel{+3.0}{_{-2.6}}$  & deg  &   RM \\
              &  $i_\mathrm{\star}^{\rm South}$   &  171.9$\stackrel{+2.5}{_{-3.4}}$  & deg   &   RM \\
Equatorial velocity    &  $v_\mathrm{eq}$    & [65.28 - 120]  &   km\,s$^{-1}$   &   RM   \\
    \hline
    \hline 
\textit{Planetary properties}    &   &   &  \\ 
    \hline
Transit epoch	& $T_{0}$ & 2458119.72074$\pm$0.00017 &   BJD$_\mathrm{TDB}$ & TESS  \\    
Orbital period & $P$ & 1.27492504$^{+1.5\times 10^{-7}}_{-1.4\times 10^{-7}}$  &  d & (TESS+EulerCam)  \\  
Scaled semi-major axis & $a_\mathrm{p}/R_{\star}$ & 3.8131$^{+0.0075}_{-0.0060}$ & & (TESS+EulerCam)  \\
Semi-major axis & $a_\mathrm{p}$ & 0.02596$^{+0.00043}_{-0.00063}$ &  au & (TESS+EulerCam)$^{\dagger}$  \\			
Eccentricity & $e$ & [0 - 0.0032]  & & TESS  \\
Argument of periastron & $\omega$ & 10$\pm$10 &   deg  &   TESS  \\
Orbital inclination 	& $i_\mathrm{p}$ & 88.49$\pm$0.16 &  deg & (TESS+RM)  \\     
Impact parameter & $b$ & 0.10$\pm$0.01 &   &   (TESS+RM)$^{\dagger}$    \\ 
Transit durations & $T_\mathrm{14}$ &  2.9053$\stackrel{+0.0065}{_{-0.0059}}$   &   h  &  TESS$^{\dagger}$ \\
				& $T_\mathrm{23}$ &   2.2605$\stackrel{+0.0055}{_{-0.0053}}$    &   h &  TESS$^{\dagger}$ \\
Planet-to-star radius ratio & $R^\mathrm{T}_\mathrm{p}/R_{\star}$ & 0.12355$\stackrel{+0.00033}{_{-0.00029}}$ & & TESS   
\\ 
 						   & $R^\mathrm{E}_\mathrm{p}/R_{\star}$ & 0.12534$\stackrel{+0.00043}{_{-0.00060}}$ &    & EulerCam
\\ 
Radius & $R^\mathrm{T}_\mathrm{p}$ &  1.753$\pm$0.036 & $R_\mathrm{Jup}$  &  TESS$^{\dagger}$      \\
       & $R^\mathrm{E}_\mathrm{p}$ &  1.773$\stackrel{+0.041}{_{-0.033}}$ & $R_\mathrm{Jup}$  &  EulerCam$^{\dagger}$      \\
Stellar reflex velocity	& $K$ & 177.0$\stackrel{+8.5}{_{-8.1}}$ &   m\,s$^{-1}$  &   RV  \\
Mass  & $M_\mathrm{p}$ &   1.157$\pm$0.070 & $M_\mathrm{Jup}$  &  RV$^{\dagger}$     \\
Density  & $\rho_\mathrm{p}$ & 0.266$\stackrel{+0.024}{_{-0.022}}$ &  g\,cm$^{-3}$  &   (TESS+RV)$^{\dagger}$   \\
Surface gravity  & $g_\mathrm{p}$ &  9.33$\stackrel{+0.71}{_{-0.67}}$  &   m\,s$^{-2}$  &   (TESS+RV)$^{\dagger}$  \\
Sky-projected obliquity    & $\lambda$  &  87.20$\stackrel{+0.41}{_{-0.45}}$  &    deg   &   RM  \\
3D obliquity    & $\psi^{\rm North}$  &  88.1$\pm$0.25  &    deg    &   RM$^{\dagger}$\\
			    & $\psi^{\rm South}$  &  91.11$\pm$0.20  &    deg    &   RM$^{\dagger}$\\	 			       
    \hline
  \end{tabular}
  \begin{tablenotes}[para,flushleft]
  Notes: Values in square brackets indicate the 1$\sigma$ confidence intervals for the equatorial velocity and eccentricity, whose probability distributions peak at the lower boundary values for these parameters. The 3$\sigma$ confidence intervals for these parameters are [65.28 - 295]\,km\,s$^{-1}$ and [0 - 0.0078]. Properties with TESS origin are reported from \citet{Bourrier2019}, or revised when combined with other datasets. Coefficients $u_1$ and $u_2$ are associated with a quadratic limb-darkening law. The daggers indicate derived parameters. Planetary density and surface gravity were calculated using the lowest planet-to-star radius ratio (from TESS). There are two possible solutions for the stellar inclination and 3D obliquity of WASP-121b, marked as \textit{North} or \textit{South} depending on which pole of the star is visible. \\
  \end{tablenotes}
  \end{threeparttable}
\label{tab:sys_prop}
\end{table*}

\section{Reloaded Rossiter-McLaughlin analysis}
\label{sec:reloaded RM}

\subsection{HARPS observations of WASP-121}
\label{sec:HARPS_data}

We studied the orbital architecture of WASP-121b and the properties of its host star by analyzing three transit observations obtained with the HARPS echelle spectrograph (HEARTS survey, ESO program 100.C-0750; PI: D. Ehrenreich). Three visits were scheduled on 31 December 2017 (Visit 1), 9 January 2018 (Visit 2), and 14 January 2018 (Visit 3). They lasted between 6.6 and 8.1\,h, covering the full duration of the transit ($\sim$2.9\,h) with sufficient baseline on each side to determine the unocculted stellar properties (Table~\ref{tab:log}).

Observations were reduced with the HARPS (version 3.8) Data Reduction Software, yielding spectra with resolving power 115,000 and covering the region 380-690 nm. The reduction includes a correction of the color effect due to variability of the extinction caused by Earth's atmosphere during the transit (e.g., \citealt{bourrier2014b}, \citealt{Bourrier_2018_Nat}). The spectrum of a hot F-type star such as WASP-121 contains far fewer absorption lines than later-type stars. Including these absent lines into the mask would reduce the contrast of the CCF and the precision on their derived properties. Furthermore, the fast rotation of WASP-121 broadens the stellar lines, blending lines that are isolated in the spectrum of colder stars. A single mask line needs to be associated with unresolved stellar lines contributing to the same blended line to avoid introducing correlated information into the CCF. We thus computed CCFs for each spectral order using a custom mask specific to WASP-121 (this mask is available in electronic form at the CDS). All measured 1D spectra were averaged and smoothed with a 0.09\,\AA\, moving average. The continuum was estimated by running a local maximum detection algorithm on the spectrum, using an alpha-shape algorithm to remove unreliable local maxima and applying a cubic interpolation on the remaining maxima to extrapolate the continuum on the full wavelength grid. The stellar lines to be included in our custom mask were then defined as a local minimum surrounded by two local maxima. A first estimate of the detectable lines and their position was made by running a local minimum detection algorithm on the stellar spectrum. Positions were then derived more accurately as the minimum of a parabola fit around the line minimum in a window of $\pm$3\,km\,s$^{-1}$. We discarded lines with windows smaller than 5 pixels, lines with derived centers farther away than 0.03\,\AA\, from the local minimum, and shallow lines with relative flux difference between the local minimum and highest local maxima smaller than 0.05. Last, we generated a synthetic telluric spectrum with Molecfit \citep{Smette2015}, and removed mask lines for which the core of a neighboring telluric line (with depth ratios with the mask line higher than 2\%) entered the region defined by the two local maxima of the mask line for at least one Earth barycentric RV of the spectrum. The final mask is composed of 1828 lines. Their weights were set to the relative flux difference between the stellar lines local minimum and the average of their two local maxima (\citealt{pepe2002}). 

Because the CCFs generated with the HARPS DRS are oversampled with a step of 0.25\,km\,s$^{-1}$, and have a pixel width of about 0.8\,km\,s$^{-1}$, we kept one in four points in all CCFs prior to their analysis (\citealt{Cegla2017}). Here we note that by construction, our custom mask lines are at rest in the stellar rest frame. The null velocity in the CCFs calculated with this mask thus directly corresponds to the stellar rest velocity. \\

\begin{table}   
\caption[]{Log of WASP-121 b HARPS transit observations}
\centering
\begin{threeparttable}
\begin{tabular}{c|c|c|c}
    \hline
    \hline
    Visits   &    1   &   2   &   3   \\
    \hline
    Date (start)  & 31-12-17 &  09-01-18 &   14-01-18 \\
    Number of exposures & 35   &  55   &  50  \\
    Exposure duration (s)  & 570-720   & 500-600     &  500-660  \\    
    Exposure S/N (550\,nm)  &  26-61  &  21-43   &  32-49  \\
    \hline
  \end{tabular}
  \begin{tablenotes}[para,flushleft]
  Notes: The S/N is given per pixel.\\
  \end{tablenotes}
  \end{threeparttable}
\label{tab:log}
\end{table}


\subsection{Simultaneous EulerCam photometry}
\label{sec:LC_fit}

\begin{center}
\begin{figure}
\centering
\includegraphics[trim=0.cm 2.cm 0.cm 0cm,clip=true,width=\columnwidth]{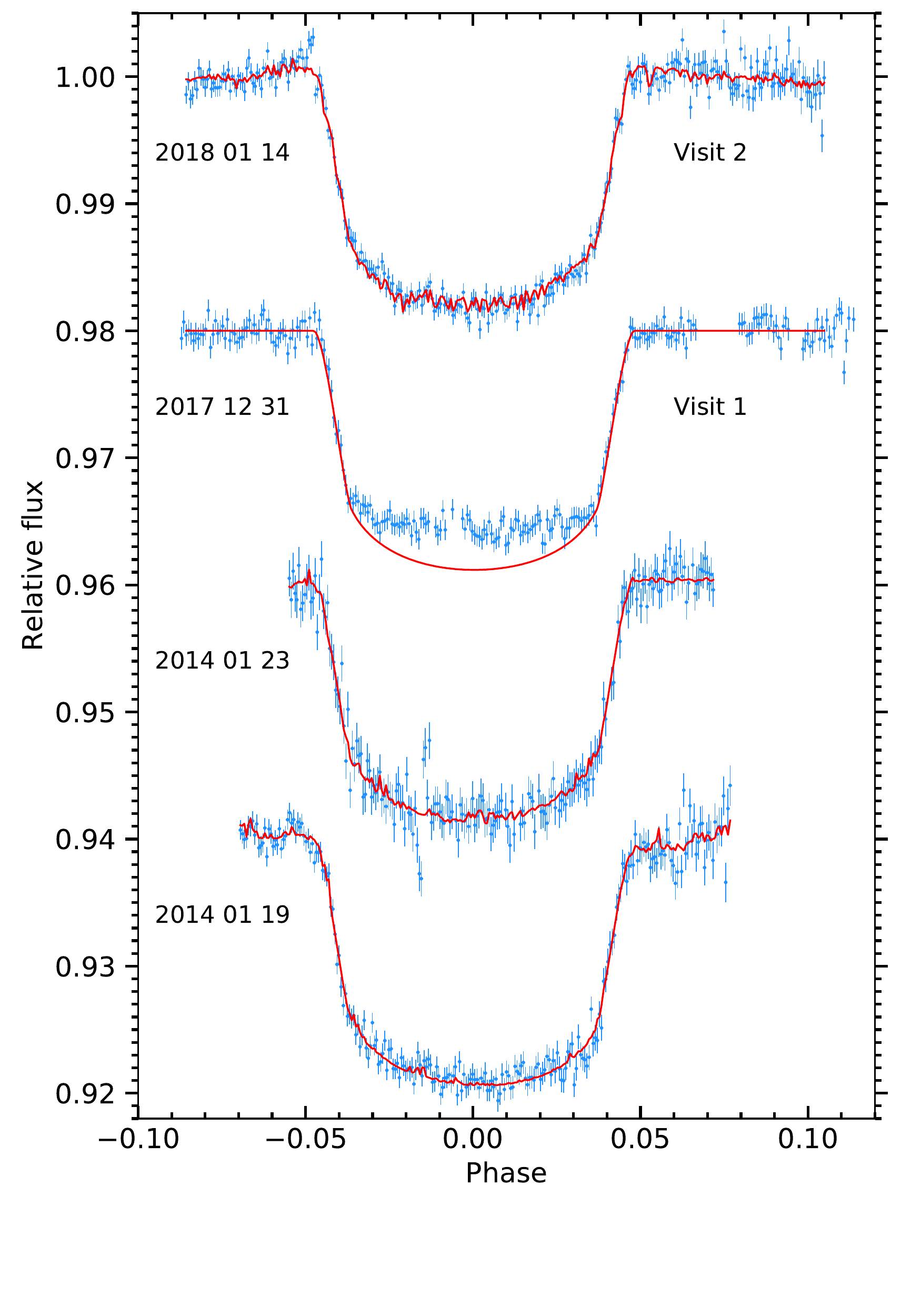}
\caption[]{Transit light curves of WASP-121b obtained with EulerCam, offset by 0.02 for visibility. Best-fit models fitted to the 2014 and 2018 data are shown in red. They include a common transit model and a detrending model specific to each visit. The abnormal shape of the 2017 light curve is likely due to instrumental effects. For this epoch we only overplot the transit model.}
\label{fig:EULER_LC_all}
\end{figure}
\end{center}

The reloaded RM technique requires knowledge of the transit light curve in the spectral band of the CCFs (\citealt{Cegla2016}). Measuring the transit simultaneously in photometry and spectroscopy further allows us to determine occulted spots and plages along the chord that is transited by the planet. We therefore obtained simultaneous photometry throughout the transits in Visits 1 and 2 using EulerCam at the 1.2m Euler telescope at La Silla. Observations were carried out using an r'-Gunn filter to match the HARPS wavelength band as closely as possible, and we applied a slight defocus to the telescope to improve the target point spread function (PSF) sampling and the observation efficiency. After standard image correction procedures, we extracted relative aperture photometry, iteratively selecting a set of stable reference stars and testing a number of extraction apertures. For details on EulerCam and the relevant observation and data reduction procedures, see \citet{Lendl2012}. We combined the new broadband photometry with two archival EulerCam light curves observed in r' band on 2014 January 19 and 23 (Fig.~\ref{fig:EULER_LC_all}, \citealt{Delrez2016}). The four light curves are shown in Fig.~\ref{fig:EULER_LC_all}, and they are available in electronic form at the CDS. 

The increased scatter during the transit on 2014 January 23 is caused by the passage of a cloud. The light curve obtained in Visit 1 shows a much shallower transit than the light curves that were obtained during the other epochs. We did not find any large variation (beyond the mmag level) in the overall stellar brightness between Visits 1 and 2, as are created, for example, by changing star spot coverage, which would translate into an offset in the measured transit depth. No variations in transit depth similar to that of Visit 1 are found in any of the 17 transits observed with TESS (\citealt{Bourrier2019}). We lack a convincing physical explanation for this anomaly, and suggest that instrumental effects linked to image saturation are likely the origin of this variation. Indeed, the target saturated the detector near the transit center in Visit 1, and the data are therefore likely affected by detector nonlinearity at high flux levels. The light curve from Visit 1 was therefore excluded from further analyses. \\

We made use of the MCMC code described in \citet{Lendl2017} to fit the EulerCam data. We assumed a uniform prior distribution for the star-to-planet radius ratio (i.e., this parameter was fit without any a priori constraints), and placed normal prior distributions on the impact parameter, the transit duration, the mid-transit time, and the planetary period. These priors were centered on the values derived in \citet{Bourrier2019}, and their width corresponds to the respective 1\,$\sigma$ uncertainties. We used the routines of \citet{Espinoza2015} to compute quadratic limb-darkening coefficients, using a wide ($\sigma_{prior}=0.1$) normal prior distribution centered on these values in our analysis. Our code allows for the use of parametric baseline models (see, e.g., \citealt{Gillon2010}), and we find that the light curves of 2014 January 19 and 23, and 2018 January 14 are best fit by models of the form $p(t^2)+p(\mathit{xy}^1) + p(\mathit{FWHM}^1)$, $p(t^1)+ p(\mathit{FWHM}^2)$, and $p(t^2)+p(\mathit{xy}^1)+p(\mathit{FWHM}^2)$, respectively, where $p(\mathit{i}^n)$ refers to a polynomial of order $n$ in parameter $i$. The parameters are the time $t$, coordinate shifts $xy$, and stellar $\mathit{FWHM}$. System properties specific to the EulerCam passband are given in Table~\ref{tab:sys_prop}. The baseline between the EulerCam observations is long, as transits are separated by several years, and therefore these transits improved the precision on the orbital period and semi-major axis compared to the TESS fit. We updated their values accordingly.\\

We compared our measurement for R$_\mathrm{p}/$R$_{*}$ (0.1253$\stackrel{+0.0004}{_{-0.0006}}$ from EulerCam in 619-710\,\AA\,) and that of \citet{Bourrier2019} (0.1236$\pm{0.0003}$ from TESS in 600-1000\,\AA\,) with those of \citet{Evans2018} obtained with the G430 and G750 HST/WFC3 grisms, averaging their measurements within the EulerCam and TESS respective passbands. The \citet{Evans2018} results yield R$_\mathrm{p}/$R$_{*}$ = 0.12238$\pm$0.00036 (EulerCam) and 0.12244$\pm$0.00021 (TESS), which is significantly lower than the EulerCam and TESS measurements by 0.003 (5.3$\sigma$) and 0.001 (2.7$\sigma$), respectively. \citet{Evans2018} previously noted that their measurements were lower than values obtained using ground-based photometry in the $B$, $r'$, and $z'$ bandpass (\citealt{Delrez2016}) and proposed that these discrepancies could arise from systematics in the latter measurements. Interestingly, our planet-to-star radius ratios are consistent with those obtained by \citet{Delrez2016} in the bands that overlap with those of EulerCam (0.12521$\pm$0.0007 in 555-670\,\AA\,) and TESS (0.12298$\pm$0.0012 in 836-943\,\AA\,). The good agreement between ground- and space-based measurements might suggest that the reduction procedure or systematics specific to the HST data might have offset the transit depths derived by \citet{Evans2018}.\\

\subsection{Analysis of the local stellar CCFs}
\label{sec:extra}

The HARPS CCFs (heareafter CCF$_\mathrm{DI}$) originate from starlight integrated over the disk of WASP-121. We used the reloaded RM technique (\citealt{Cegla2016}, see also \citealt{Bourrier2017_WASP8,Bourrier_2018_Nat}) to isolate the local CCF (hereafter CCF$_\mathrm{loc}$) from the regions of the photosphere that are occulted by WASP-121b. The CCF$_\mathrm{DI}$ calculated in the stellar rest frame were first corrected for the stellar Keplerian motion induced by WASP-121b. We identified the CCF$_\mathrm{DI}$ obtained outside of the transit, taking care to exclude those that even partially overlapped with the transit window, and coadded them to build a ``master-out'' CCF$_\mathrm{DI}$ in each night, which corresponds to the unocculted star. The continua of the master-out and individual CCF$_\mathrm{DI}$ outside of the transit were normalized to the same continuum at unity, while in-transit CCF$_\mathrm{DI}$ were scaled to reflect the planetary disk absorption. This scaling was made using the theoretical transit light curve derived from the fit to the EulerCam data (Sect.~\ref{sec:LC_fit}), whose spectral range is closer to that of HARPS than that of TESS. 

\begin{center}
\begin{figure}[tbh!]
\centering
\includegraphics[trim=0cm 0cm 0cm 0cm,clip=true,width=\columnwidth]{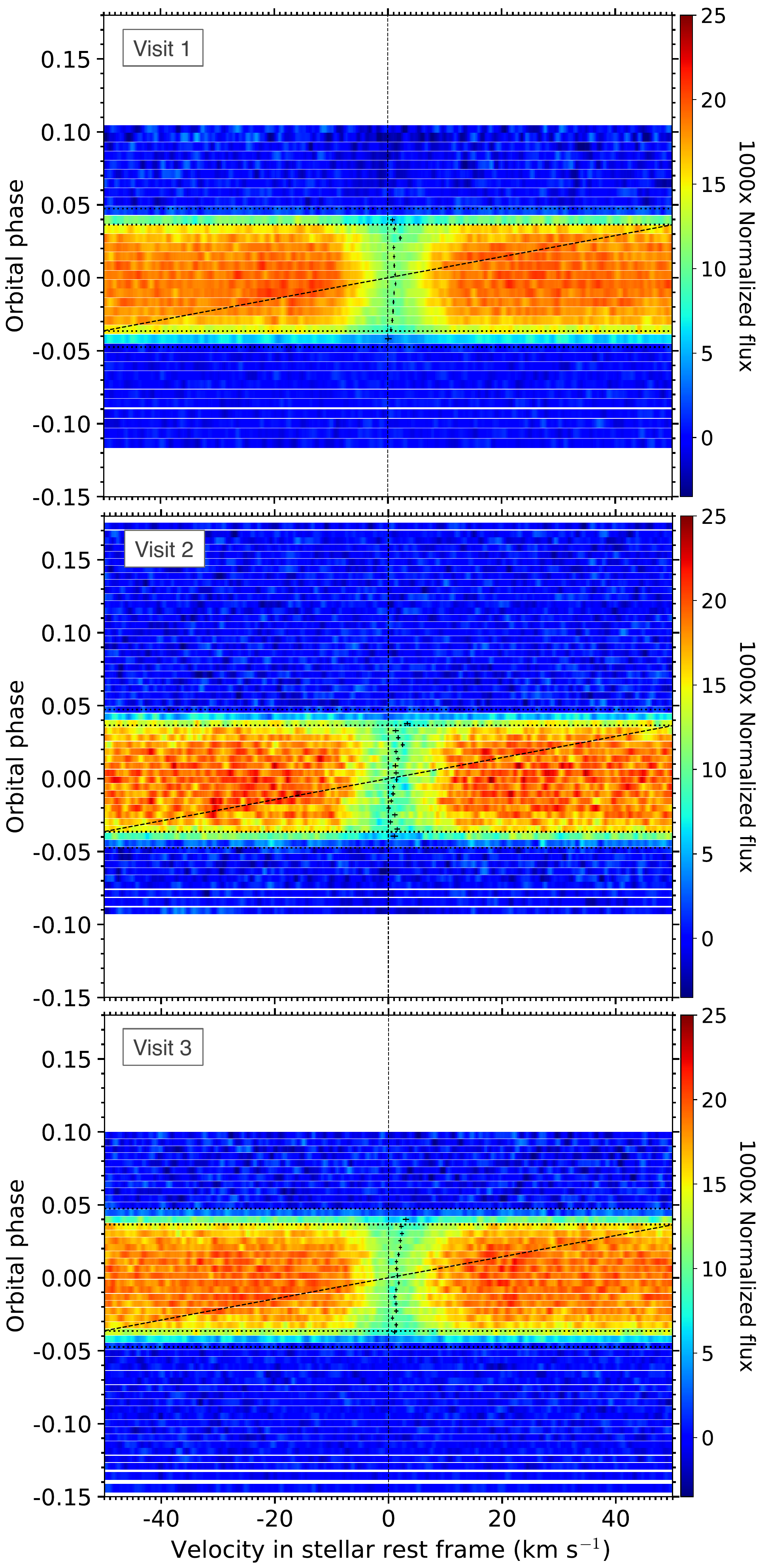}
\caption[]{Maps of the residuals between the scaled CCF$_\mathrm{DI}$ and their master-out in each visit. Residuals are colored as a function of their flux, and plotted as a function of RV in the stellar rest frame (in abscissa) and orbital phase (in ordinate). The vertical dashed black line indicates the stellar rest velocity. Horizontal dotted lines are the transit contacts. In-transit residuals correspond to the CCF$_\mathrm{loc}$, and show the average local stellar line profile (recognizable by a lower flux in the CCF$_\mathrm{loc}$ cores) from the planet-occulted regions of the stellar disk. For comparison, the spectroscopic width of the disk-integrated stellar lines is about 14\,km\,s$^{-1}$. Black crosses with error bars indicate the centroids of the detected stellar line profile. The slanted dashed black line tracks the orbital trajectory of the planet.}
\label{fig:2D_maps}
\end{figure}
\end{center}

The CCF$_\mathrm{loc}$ associated with the planet-occulted regions were retrieved by subtracting the scaled in-transit CCF$_\mathrm{DI}$ from the master-out in each night. The local stellar line profiles from the planet-occulted regions of the photosphere are clearly visible in Fig.~\ref{fig:2D_maps}. They are always redshifted, and this redshift slightly increases along the transit chord. WASP-121b therefore always transits that hemisphere of the star that rotates away from us, with a transit chord farther from the projected stellar spin axis at egress than at ingress. This preliminary analysis implies that the sky-projected obliquity $\lambda$ must be slightly lower than 90$^{\circ}$, in contrast to the value of 102.2$\pm$5.5$^{\circ}$ (using the same convention as in the present study) derived by \citealt{Delrez2016} from a classical velocimetric analysis of the RM effect in CORALIE data.

\begin{center}
\begin{figure}
\centering
\includegraphics[trim=0cm 0cm 0cm 0cm,clip=true,width=\columnwidth]{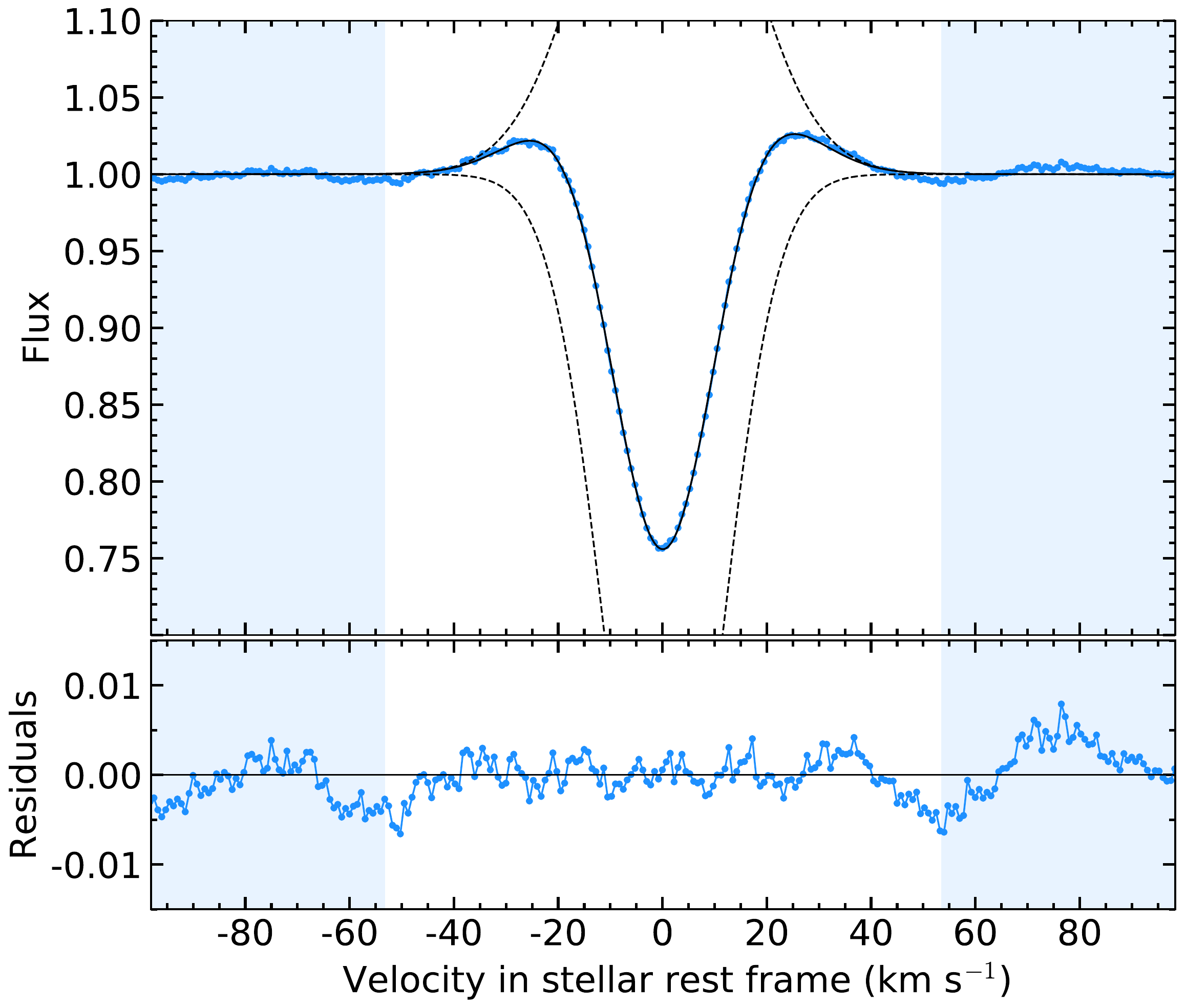}
\caption[]{Typical CCF$_\mathrm{DI}$ integrated over the disk of WASP-121 (blue points, obtained during one of the out-of-transit exposures in Visit 2). The solid black profile is the best-fit double-Gaussian model to the measured CCF. The dashed black profiles show the individual Gaussian components to this model, which yields a low dispersion on the fit residual (bottom panel). The blue shaded regions indicate the velocity ranges used to define the CCF continuum.}
\label{fig:CCF_DI_fit}
\end{figure}
\end{center}

The RV centroids of the CCF$_\mathrm{loc}$ can generally be derived from a simple Gaussian fit. The CCFs generated with our custom mask for WASP-121, however, show side lobes that would limit the precision of CCF properties derived with a Gaussian fit. Therefore, we used the double-Gaussian model introduced by \citet{Bourrier_2018_Nat} for the M dwarf GJ\,436, which consists of the sum of a Gaussian function representing the CCF continuum and side lobes, and an inverted Gaussian function representing the CCF core. As illustrated in Fig.~\ref{fig:CCF_DI_fit}, the double-Gaussian model reproduces the entire CCF of WASP-121 well and thus exploits the full information contained in its profile. 

\begin{center}
\begin{figure}
\centering
\includegraphics[trim=1.5cm 0cm 1cm 0cm,clip=true,width=\columnwidth]{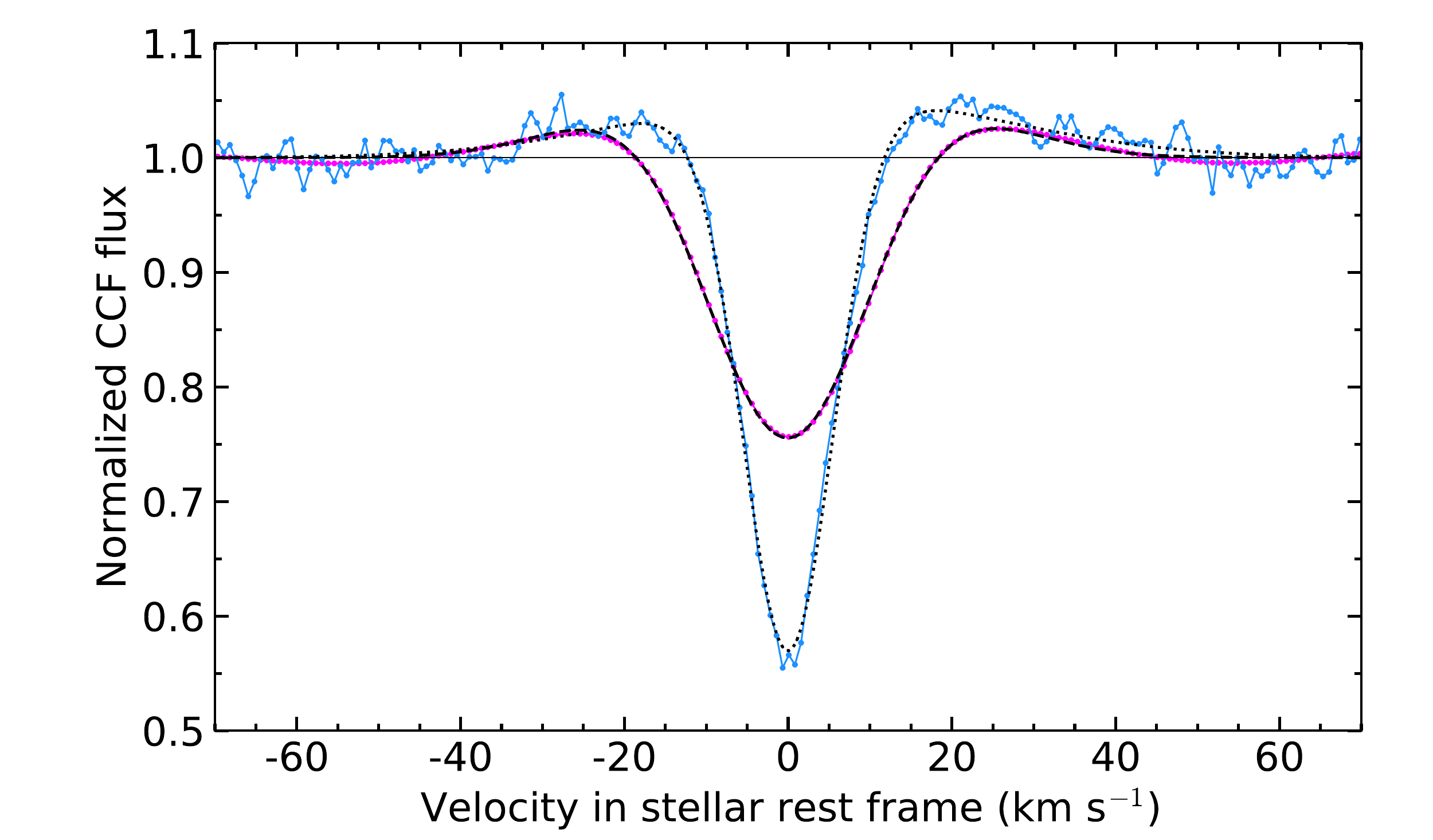}
\caption[]{Master-out CCF$_\mathrm{DI}$ (magenta) and master local CCF$_\mathrm{loc}$ (blue), binned over the three visits and normalized to the same continuum. The dashed and dotted black profiles show the best-fit models to the master-out and master-local, respectively. They are based on the same double-Gaussian model, but with different correlations between the properties of the lobe and core Gaussian components.}
\label{fig:Compa_Models_Out_Loc}
\end{figure}
\end{center}

We performed a preliminary fit to the CCF$_\mathrm{loc}$ using a double-Gaussian model where the FWHM ratio, contrast ratio, and centroid difference between the core and lobe components were set to the best-fit values for the nightly master-out CCF$_\mathrm{DI}$ (as in \citealt{Bourrier_2018_Nat}). The local average stellar line is considered detected if the amplitude of the model CCF$_\mathrm{loc}$ (defined as the flux difference between the minimum of the model CCF$_\mathrm{loc}$ and its continuum) is three times larger than the dispersion in the continuum of the observed CCF$_\mathrm{loc}$. This led us to discard a few CCF$_\mathrm{loc}$ that were located very near the stellar limb, where the lower flux and partial occultation by the planet yield very low S/Ns ratios. The remaining CCF$_\mathrm{loc}$ were shifted to the same rest velocity and averaged on each night to create a master local CCF$_\mathrm{loc}$ (e.g., \citealt{Wyttenbach2017}). The comparison with the master CCF$_\mathrm{DI}$ (Fig.~\ref{fig:Compa_Models_Out_Loc}) clearly shows the effect of rotational broadening; the local average stellar line is far narrower and deeper than the disk-integrated line. Both CCFs show sidelobes, which are well fit with a double-Gaussian model but with different properties. In the master CCF$_\mathrm{loc}$ the lobe component is broader, and more redshifted relative to the core component, than in the master CCF$_\mathrm{DI}$. The final fit to the CCF$_\mathrm{loc}$ in individual exposures was performed with a double-Gaussian model where the core and lobe components were linked as in the nightly master CCF$_\mathrm{loc}$. Flux errors assigned to the CCF$_\mathrm{loc}$ were set to the standard deviation in their continuum flux, and the uncertainties on the derived parameters were set to the 1\,$\sigma$ statistical errors from a Levenberg-Marquardt least-squares minimisation. The local stellar surface RVs were defined as the derived centroids of the CCF$_\mathrm{loc}$ core component.\\

\subsection{Analysis of the stellar rotation and orbital architecture}
\label{sec:fit_RM}

\subsubsection{Model and prior constraints}
\label{sec:priors_RM}

\begin{center}
\begin{figure}
\centering
\includegraphics[trim=0cm 0.cm 0cm 0cm,clip=true,width=\columnwidth]{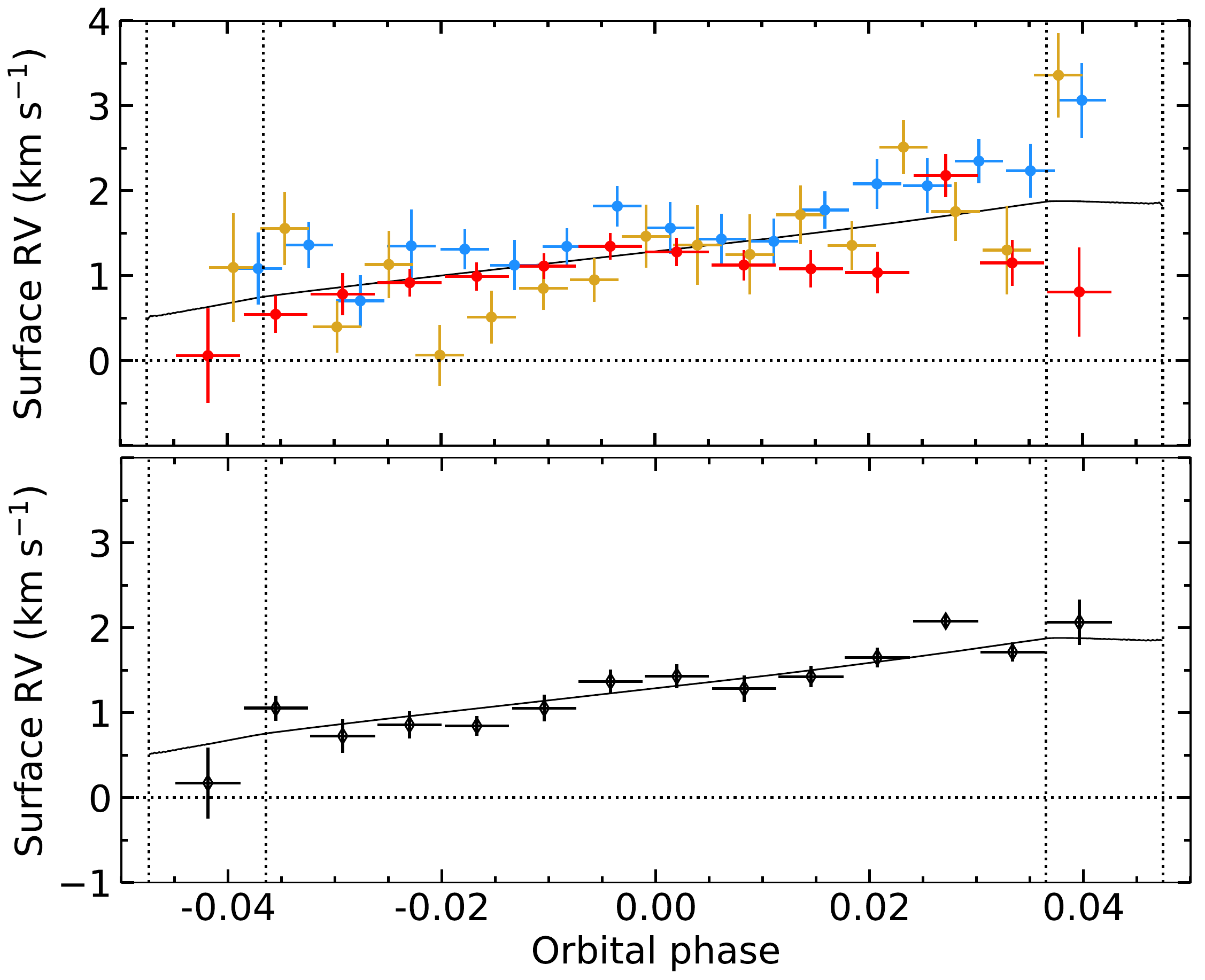}
\caption[]{Radial velocities of the stellar surface regions occulted by WASP-121b as a function of orbital phase. Horizontal bars show the exposure durations. The black curve is the best-fit reloaded RM model (indistinguishable between the low- and high- $i_{*}$ solutions) to the three visits. Dashed vertical lines are the transit contacts. The horizontal dashed line highlights the stellar rest velocity, which is found along the projected stellar spin axis. \textbf{Upper panel}: Local RVs in individual Visits 1 (red), 2 (gold), and 3 (blue). \textbf{Bottom panel}: Local RVs derived from the CCF$_\mathrm{loc}$ binned over the three visits, shown separately for the sake of clarity.}
\label{fig:RV_local}
\end{figure}
\end{center}

Despite some variability, the local RVs follow a similar trend in the three visits (Fig.~\ref{fig:RV_local}). They become more redshifted along the transit chord and remain always positive, confirming the preliminary interpretation performed in Sect.~\ref{sec:extra} of a near-polar orbit only crossing the redshifted stellar hemisphere. The orbital architecture of the system and the properties of the velocity field of the stellar photosphere can be derived from the fit to the local RVs using the reloaded RM model (\citealt{Cegla2016}; see their Figure 3 for the definitions of the coordinate system and angle conventions), which calculates brightness-weighted theoretical RVs averaged over each planet-occulted region. In previous reloaded RM studies (\citealt{Cegla2016}, \citealt{Bourrier2017_WASP8, Bourrier_2018_Nat}) the model was fit to the data by varying the stellar projected rotational velocity $v_{\rm eq}\sin i_{*}$ (and in some cases, the differential rotation or convective motions of the stellar photosphere) and the sky-projected obliquity $\lambda$. The latter parameter alone thus controlled the model planet trajectory, and the coordinates of the occulted regions. The near-polar orbit of WASP-121b, however, results in $v_{\rm eq}\sin i_{*}$ being strongly degenerate with the planet impact parameter (see Appendix~\ref{apn:polar_orb}), which remains poorly determined because of the uncertainty on the orbital inclination $i_\mathrm{p}$ (\citealt{Bourrier2019}). Interestingly, an impact parameter close to zero would require that the planet cross the projected stellar spin axis (where local RVs are zero), which is incompatible with the transit of a single stellar hemisphere indicated by the positive local RVs series. This means that more stringent constraints can be derived on the orbital inclination from the fit to the local RVs, and we therefore modified the reloaded RM model to include $i_\mathrm{p}$ as a free parameter. The scaled semi-major axis of WASP-121b has less influence on the local RVs and is much better determined than $i_\mathrm{p}$. After checking that $a_\mathrm{p}/R_{*}$ could not be better constrained through the fit, we therefore fixed it to its nominal value. Similarly, the other orbital properties and the ephemeris of WASP-121b are known to a much higher precision through transit photometry and velocimetry than could be obtained via the fit to the local RVs, and they were accordingly fixed to their nominal values. The planet-to-star radius ratio and the stellar limb-darkening coefficients cannot be retrieved from the fit to the local RVs because absolute flux levels are lost in HARPS ground-based data. We note that the measured local RVs do not depend on our choice for $i_\mathrm{p}$ because the photometric scaling of the CCFs was performed directly with the transit light-curve model fit to the simultaneous EulerCam data (Sect.~\ref{sec:extra}). \\

\begin{center}
\begin{figure}
\centering
\includegraphics[trim=1cm 0cm 1cm 0.5cm,clip=true,width=\columnwidth]{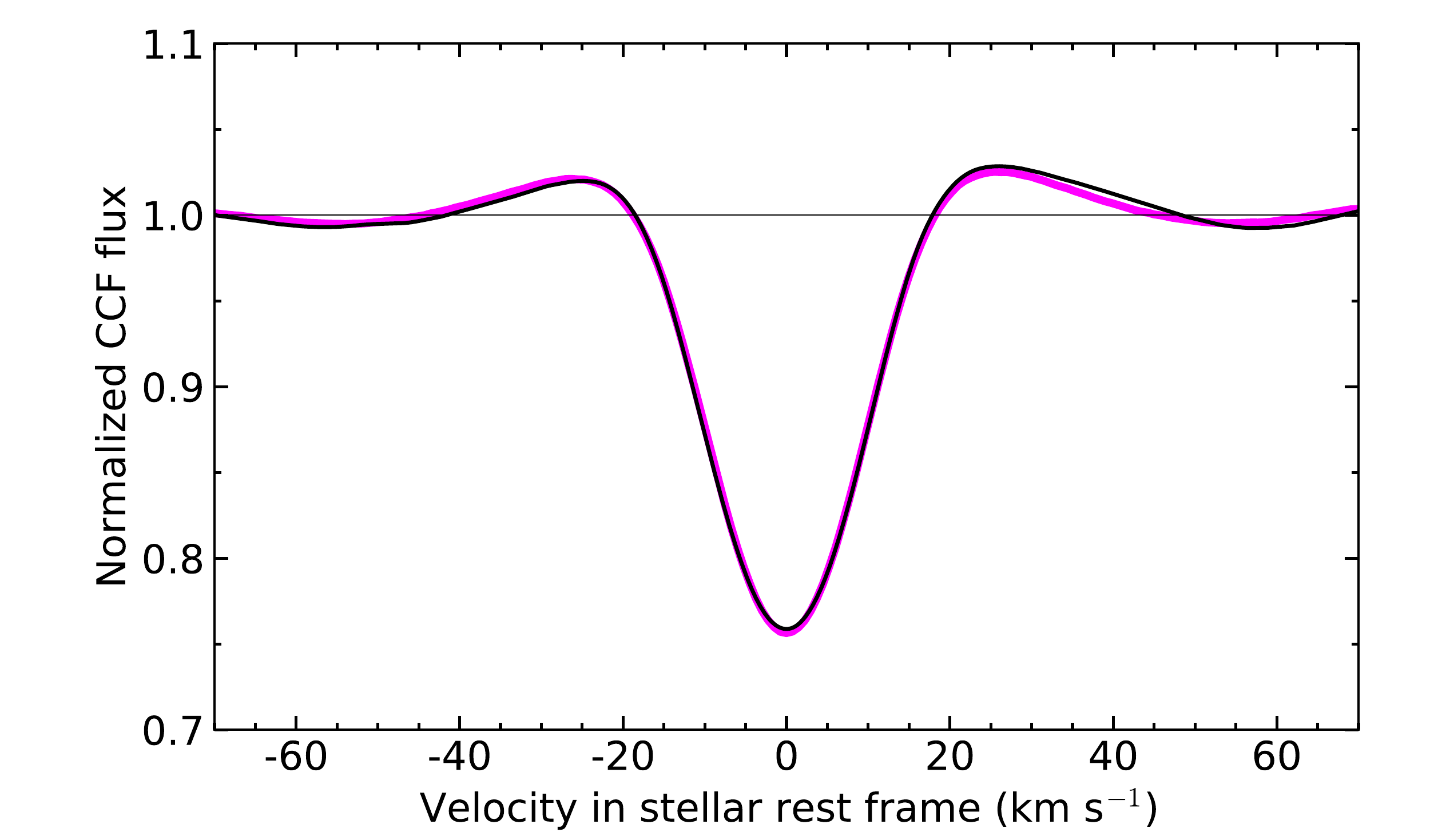}
\caption[]{Master-out CCF$_\mathrm{DI}$ (thick magenta line) and its best fit (black line) obtained by tiling a model star with the limb-darkened master local CCF$_\mathrm{loc}$, used as a proxy for the specific stellar intensity profile. The best-fit value for the projected stellar rotational velocity, which controls the spectral position of this profile over the model stellar disk, is in good agreement with its measured spectroscopic value and yields a good fit to the observed master-out CCF$_\mathrm{DI}$.}
\label{fig:Fit_Mout_Mloc}
\end{figure}
\end{center}

Additional constraints can be set from the independent measurements of stellar line broadening and the stellar rotational period. \citet{Delrez2016} derived a spectroscopic value $v_{\rm eq}\sin i_{*/\rm spec}$ = 13.56$\stackrel{+0.69}{_{-0.68}}$\,km\,s$^{-1}$ from the fit to stellar \ion{Fe}{i} lines in CORALIE spectra. A similar estimate can be derived from the comparison between the HARPS master-out CCF$_\mathrm{DI}$ and master-local CCF$_\mathrm{loc}$. Under the assumption that CCF$_\mathrm{loc}$ measured along the transit chord are representative of the entire stellar disk, the observed master-out was fit by tiling a model star with the master-local CCF$_\mathrm{loc}$, weighted by the limb-darkening law derived from the EulerCam photometry, and shifted in RV position by the solid rotation of the photosphere, which was let free to vary. We obtain a good fit for $v_{\rm eq}\sin i_{*} \sim $13.9\,km\,s$^{-1}$ (Fig.~\ref{fig:Fit_Mout_Mloc}), suggesting that the local average stellar line profile does not change substantially across the stellar disk within the precision of HARPS, and that $v_{\rm eq}\sin i_{*/\rm spec}$ can be used as a prior for the stellar projected rotational velocity. 

Analysis of ground-based spectroscopy and TESS photometry of WASP-121 (Sect.~\ref{sec:RV_fit}) revealed a persistent rotational modulation at 1.13 days, and a transient modulation at 1.37 days. We understand these results as an indication of differential rotation, with the equator of WASP-121 rotating at least as fast as the latitudes probed by the 1.13 days signal, and the transient signal arising from higher latitudes that rotate more slowly. This sets a prior on the stellar equatorial velocity $v_{\rm eq}\geqslant$ 65.28\,km\,s$^{-1}$.

The three local RV series were simultaneously fit with the updated RM model. We assumed a solar-like differential rotation law P$(\theta_\mathrm{lat})$ = P$_\mathrm{eq}/(1 - \alpha$\,sin$^{2}(\theta_\mathrm{lat}))$, where $\theta_\mathrm{lat}$ is the stellar latitude and $\alpha = 1 - $P$_\mathrm{eq}/$P$_\mathrm{pole}$ is the relative differential rotation rate (\citealt{Cegla2016}). We accounted for the blur caused by the planetary motion during a HARPS exposure by oversampling the transit chord between the planetary position at the beginning and end of each exposure (\citealt{Bourrier2017_WASP8}). We sampled the posterior distributions of the model parameters using \textit{emcee} MCMC (Foreman2013), as in \citet{Bourrier_2018_Nat}. Jump parameters for the MCMC are the stellar equatorial velocity $v_{\rm eq}$, the cosine of the stellar inclination cos$(i_{*})$, the sky-projected obliquity $\lambda$, the orbital inclination $i_\mathrm{p}$, and the differential rotation rate $\alpha$. We set a uniform prior on $v_{\rm eq}$ and a Gaussian prior on $v_{\rm eq}\sin i_{*/\rm spec}$, following the above discussion. The posterior distribution from the fit to TESS photometry was used as prior for $i_\mathrm{p}$ (\citealt{Bourrier2019}). Uniform priors were set on the other parameters over their definition range: [-1 ; 1] for cos$(i_{*})$, [-180 ; 180]$^{\circ}$ for $\lambda$, and [-1 ; 1] for $\alpha$. \\

\subsubsection{Results}
\label{sec:results_RM}

\begin{figure}
\centering
\includegraphics[trim=0.5cm 3cm 0.5cm 1cm,clip=true,width=\columnwidth]{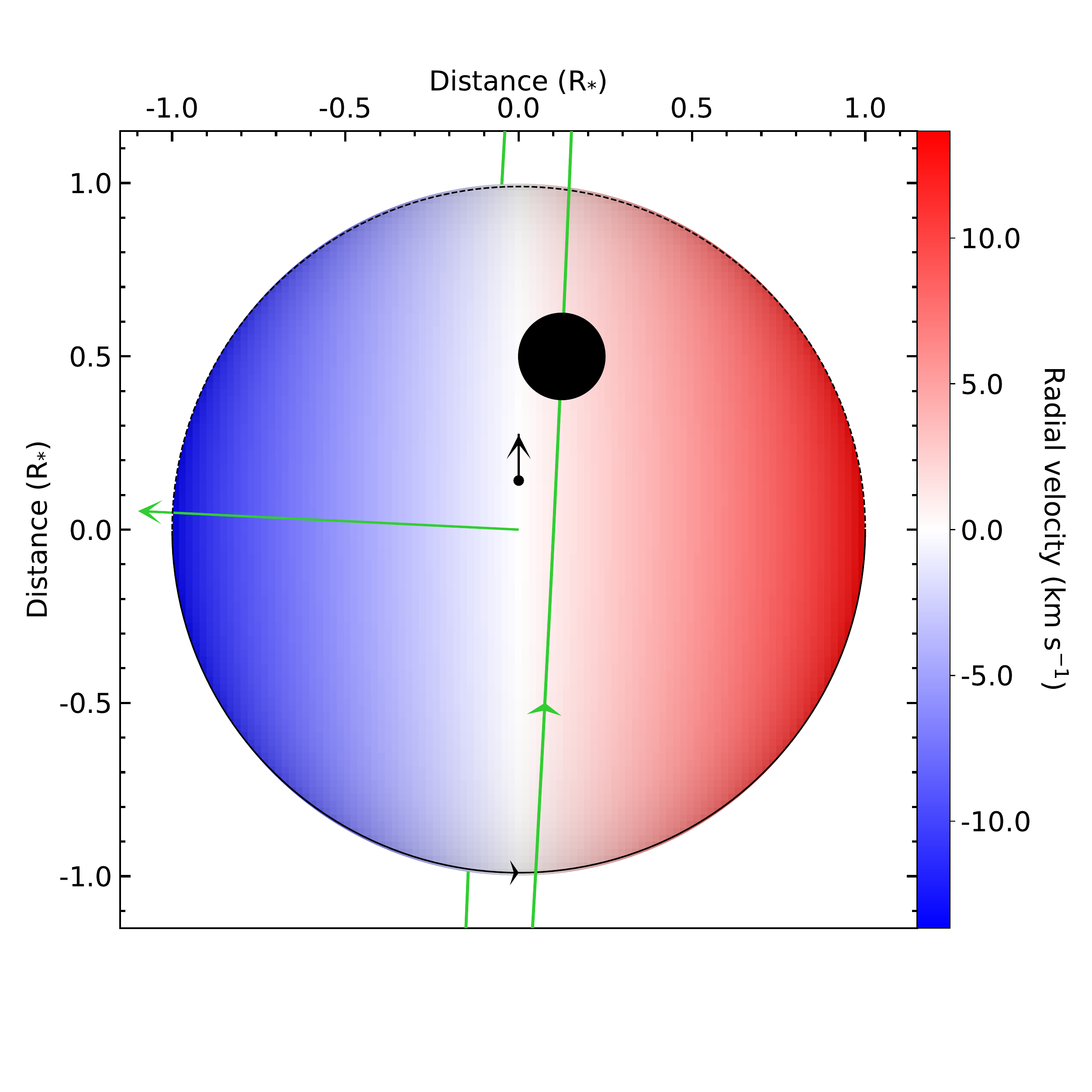}
\caption[]{Projection of WASP-121 in the plane of sky for the best-fit scenario where the north pole of the star is visible. The stellar spin axis is displayed as a black arrow extending from the north pole. The stellar equator is represented as a black line, solid when visible and dashed when hidden from view. The stellar disk is colored as a function of its RV field. The normal to the orbital plane is shown as a green arrow, and the orbital trajectory is displayed as a green curve. The black disk is WASP-121b, to scale.}
\label{fig:disque}
\end{figure}

Posterior probability distributions are shown in Figs.~\ref{fig:PD_lowi} and \ref{fig:PD_highi}. Best-fit values for the model parameters were set to the median of their distributions, and are given in Table~\ref{tab:sys_prop}. Some of the parameter distributions are asymmetrical, and we therefore chose to define their $1\sigma$ uncertainties using the highest density intervals, which contain 68.3\% of the posterior distribution mass such that no point outside the interval has a higher density than any point within it. The probability distributions show unique solutions for all model parameters, except for the stellar inclination. While we find that WASP-121 is highly inclined, the data do not allow us to distinguish whether the south pole ($i_\mathrm{*}$ = 171.9$\stackrel{+2.5}{_{-3.4}}^{\circ}$) or the north pole ($i_\mathrm{*}$ = 8.1$\stackrel{+3.0}{_{-2.6}}^{\circ}$) is visible. Both scenarios yield similar $\chi^{2}$ of 111 for 43 degrees of freedom. The relatively high reduced $\chi^{2}$ (2.6) is caused by the dispersion of the local RV measurements between the three nights. Deviations from the nominal best-fit model beyond the photon noise are present in all nights and in all phases of the transit, suggesting that variability in the local photospheric properties of this active star could be the origin of these variations. The noise in individual CCF$_\mathrm{loc}$ prevents us from searching for variations in their bisector span. No clear correlations were found between the local RVs and the FWHM or contrast of the CCF$_\mathrm{loc}$, with the EulerCam photometry, or with the Ca\,II$_\mathrm{HK}$, H$\alpha$, and Na activity indexes. We show the best-fit model for the local stellar RVs in Fig.~\ref{fig:RV_local} and the orbital architecture corresponding to the visible north pole in Fig.~\ref{fig:disque}. 

The stellar equatorial rotation remains poorly constrained, with a highest density interval of [65.28 - 120]\,km\,s$^{-1}$ that corresponds to rotation periods between [0.61 - 1.13]\,days. The probability distribution for $v_\mathrm{eq}$ nonetheless favors low velocities, which suggests that the persistent 1.13-day signal measured in photometry and ground-based data arises from active regions close to the stellar equator. We cannot confirm the differential rotation of WASP-121, with $\alpha$ = 0.08$\stackrel{+0.11}{_{-0.13}}$, but this result excludes high differential rotation rates and is consistent within 1$\sigma$ with the observed rotational modulations. Indeed, the constraints $P_\mathrm{eq}\leqslant$ 1.13\,days and $P_\mathrm{pole}\geqslant$ 1.34\,days imply $\alpha\geqslant$ 0.16. These results are also consistent with measurements obtained for Kepler stars by \citet{Balona2016}, who showed that $|\alpha|$ ranges between 0 and 0.2 for stars with rotation periods on the order of 1 day (see their Figure 9). We note that even in the case of differential rotation, the signal measured at $\sim$8.4\,days (Sect.~\ref{sec:Prot}) cannot trace the rotational modulation of a high-latitude region because the lowest $\alpha$ required would be 0.87 at the stellar poles. Measurements of the local surface RVs at higher S/N, for instance, with the ESPRESSO spectrograph, will be crucial in assessing the differential rotation of WASP-121.\\

The orbit of WASP-121b is almost but not exactly edge-on ($i_\mathrm{p}$ = 88.49$\pm$0.16$^{\circ}$) and polar ($\lambda$ = 87.20$\stackrel{+0.41}{_{-0.45}}^{\circ}$). We substantially improved the precision on these properties compared to previous studies, and find that $\lambda$ is 15$^{\circ}$ lower (3$\sigma$) than the value derived by \citet{Delrez2016} (we converted their spin-orbit angle $\beta$ = 257.8$^{\circ}$ in the same frame as our study). We combined the probability distributions of $i_\mathrm{p}$, $\lambda$, and $i_{*}$ to derive the 3D obliquity of the system, $\psi$ = arccos(sin\,$i_{*}$ cos\,$\lambda$ sin\,$i_\mathrm{p}$ + cos\,$i_{*}$ cos\,$i_\mathrm{p}$), and measure $\psi^{\rm South}$ = 91.11$\pm$0.20$^{\circ}$ (stellar south pole visible) or $\psi^{\rm North}$ = 88.1$\pm$0.25$^{\circ}$ (north pole visible). We note that our result for the obliquity does not change the conclusion by \citet{Delrez2016} that WASP-121b is on a highly misaligned orbit, and that it likely underwent strong dynamical interactions with a third companion, possibly an outer planet, during the life of the system (1.5$\pm$1.0\,Gyr). The dynamical evolution of WASP-121b is now controlled by tidal interactions with the star, leading to a gradual decrease in the obliquity and semi-major axis of the planet and to its eventual disruption (\citealt{Delrez2016}). Even with a strong tidal dissipation, however, it would take millions of years to decrease the obliquity by one degree (\citealt{Delrez2016}), and our value for the semi-major axis is not significantly lower than that of \citealt{Delrez2016}. This mechanism therefore cannot explain the difference between our measurement for $\lambda$ and that of \citet{Delrez2016}, which could be due to a bias induced by their use of the classical RM technique (\citealt{Cegla2016a}). An interesting alternative might be the nodal precession of the orbit, however, as is the case for the ultra-hot Jupiter WASP-33b (\citealt{Johnson_WASP33}). The uncertainties on the orbital inclination and obliquity from \citet{Delrez2016} prevent us from measuring a clear variation in the argument of the ascending node, with a decrease of -0.95$\pm$0.64$^{\circ}$ in about three years (from the end of 2014 to the end of 2017). Interestingly this decrease would correspond to a stellar gravitational quadrupole moment of 9.0$\times$10$^{-4}$ (for $\Psi_\mathrm{North}$) or -1.5$\times$10$^{-4}$ (for $\Psi_\mathrm{South}$), however, calculated with the equation in \citet{Barnes2013}. A negative moment is excluded by the expected oblateness of WASP-121, but the former solution is on the same order as moments estimated for the early-type fast-rotating star WASP-33 (\citealt{Johnson_WASP33,Johnson_WASP33_erratum}, \citealt{Iorio2016}).\\

\section{Distinguishing the planetary and stellar atmospheres}
\label{sec:atmo_struc}

Atmospheres of ultra-hot Jupiters were recently found to contain atomic metal species (e.g., \citealt{Hoeijmakers2018, Hoeijmakers2019}), which are prevented from condensing by the high temperatures (e.g., \citealt{Visscher2010}, \citealt{Wakeford2017}). As an ultra-hot Jupiter, WASP-121b receives extreme amounts of stellar radiation and likely undergoes atmospheric escape orders of magnitude larger than for hot Jupiters (\citealt{Salz2019_NUV_WASP121b}). Magnesium and iron ions were recently detected in its exosphere through their near-UV absorption lines (\citealt{Sing2019}), consistent with the marginally larger transit depth measured in broadband near-UV by \citet{Salz2019_NUV_WASP121b}. These metallic species likely become photoionized within the exosphere after being carried upward by the hydrodynamically expanding upper atmosphere. They could be present in their neutral form in the atmosphere of WASP-121b, and yield strong absorption in optical lines. The custom mask we built to define the CCFs of WASP-121 is based on the stellar spectral absorption lines (Sect.~\ref{sec:HARPS_data}), most of which arise from iron in the stellar atmosphere. Indeed, cross-matching our mask with the VALD database (\citealt{Piskunov1995,Kupka2000,Ryabchikova2015}) shows that of their 989 lines they have in common, more than half (570) arise from neutral iron. The second most frequent species is neutral nickel, with 67 lines. This means that if the atmospheric limb of WASP-121b contains atomic iron, we would expect its average signature to be superimposed on the stellar CCF measured during transit. We present here a summary of the technique that we devised to search for and extract the atmospheric absorption signal of an exoplanet, based on the reloaded RM approach. It will be fully described in a forthcoming paper. \\

\subsection{Method}

The in-transit CCF$_\mathrm{loc}$ extracted in Sect.~\ref{sec:extra} corresponds to the specific intensity spectrum of the occulted stellar region, multiplied by the wavelength-dependant occulting area of the planet. The intensity spectrum contains the cross-correlated absorption line from the stellar photosphere, centered at the RV of the occulted region. The occulting area is the sum of the continuum level set by the opaque planetary layers (here averaged over the HARPS band) and the equivalent surface of the atmospheric limb. If the planet contains species that absorb the CCF mask lines, this surface corresponds to the cross-correlated absorption line from the planetary atmosphere, centered at the orbital RV of the planet. When the stellar and planetary absorption lines follow sufficiently different tracks in RV-phase space, as is the case with WASP-121b (Fig.~\ref{fig:2D_maps}), it is possible to distinguish their individual contributions from the CCF$_\mathrm{loc}$.\\

\begin{figure}
\centering
\includegraphics[trim=0cm 0cm 0.cm 0cm,clip=true,width=\columnwidth]{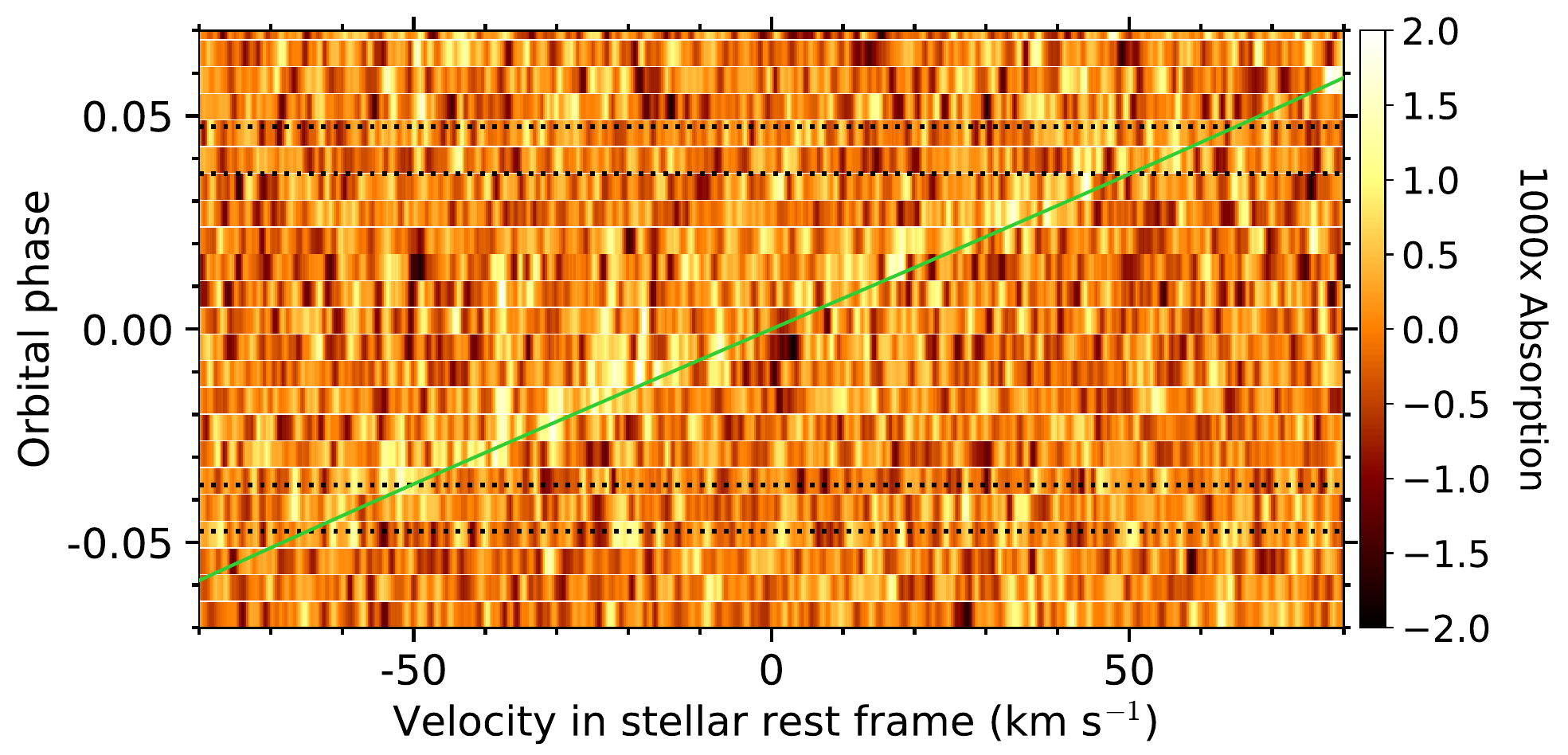}
\caption[]{Map of the atmospheric CCF$_\mathrm{atm}$ binned over the three visits, colored as a function of absorption, and plotted as a function of RV in the stellar rest frame (in abscissa) and orbital phase (in ordinate). Horizontal dotted lines are the transit contacts. The bright streak is the absorption signature from the planetary atmosphere. It follows the track of the planetary orbital motion (solid green curve), but with a slight blueshift.}
\label{fig:2D_atmo_maps}
\end{figure}

\begin{enumerate}
\item The first step consists of subtracting the stellar light that is occulted by the planetary continuum from the CCF$_\mathrm{loc}$. To do this, we used the master CCF$_\mathrm{loc}$, assuming that it is representative of the individual CCF$_\mathrm{loc}$ along the transit chord (see Sect.~\ref{sec:extra}). The master was rescaled to the correct photometric level using the best-fit EulerCam transit model (Sect.~\ref{sec:LC_fit}), which accounts for the limb-darkening and planetary continuum associated with each exposure. The rescaled master was then shifted to the RV of the planet-occulted regions, calculated with the best-fit model for the local stellar surface RVs (Sect.~\ref{sec:results_RM}). These operations yield the CCF of the product between the local stellar spectra and the transmission spectrum of the atmospheric limb in each exposure.

\item The second step consists of dividing these CCFs by the master CCF$_\mathrm{loc}$, rescaled and shifted as described in the first step, to isolate the cross-correlated absorption line of the atmospheric limb, or CCF$_\mathrm{atm}$. The scaling was made using the total surface of the star rather than the surface associated with the planetaty continuum, to obtain CCF$_\mathrm{atm}$ in classical units of absorption relative to the stellar surface. The RV-phase maps of the CCF$_\mathrm{atm}$ from WASP-121b reveal a bright streak aligned with the orbital trajectory of the planet, which is visible only during transit, and is therefore consistent with absorption by metals in the atmosphere of WASP-121b. 

\item The third and last step consists of shifting all CCF$_\mathrm{atm}$ into the planet rest frame, and averaging them over exposures where the entire planet occults the star. We calculated the theoretical RV track of the planet in the stellar rest frame using the orbital properties of the planet listed in Table~\ref{tab:sys_prop}. Ingress and egress are excluded because they probe a smaller fraction of the planetary atmosphere that varies in time. Observing WASP-121b with higher-sensitivity spectrographs such as ESPRESSO might allow studying the shape of the planetary signal during ingress/egress, and possibly resolving longitudinal variations in the planetary atmosphere. We note that we analyzed the three HARPS visits binned together, because of the small amplitude of the planetary signal, and so that the master CCF$_\mathrm{loc}$ could be determined with a high SNR. The low dispersion of residuals outside of the planetary track in Fig.~\ref{fig:2D_atmo_maps} confirms that, within the precision of the HARPS data, the master CCF$_\mathrm{loc}$ is representative of the stellar line along the transit chord. 
\end{enumerate}
The interest of this approach is that it allows us to directly use the local stellar lines that are measured along the transit chord to correct for the bias of the atmospheric signal induced by the RM effect (e.g., \citealt{Louden2015}, \citealt{Casasayas2017,Casasayas2018,Casasayas2019}). One caveat is that step 2 divides the CCF of the product between planetary and stellar lines by the CCF of the stellar lines. Unless all dominant planetary or stellar lines in the CCF mask keep the same profile, this division does not fully remove the contribution of the stellar lines in exposures where they overlap with the planetary lines. We will address this caveat in the forthcoming paper.\\

We performed a preliminary analysis to identify the velocity range that is absorbed by the planetary atmosphere in each exposure. We then carried out the reloaded RM analysis again (Sect.~\ref{sec:reloaded RM}), excluding these planet-absorbed ranges from the fits to the CCF$_\mathrm{loc}$ and from the construction of the masters CCF$_\mathrm{out}$ and CCF$_\mathrm{loc}$. In four exposures (from phase -0.007 to 0.018) the RVs of the transit chord and planetary orbit are too close to fit the uncontaminated local stellar line and retrieve its centroid (see Fig.~\ref{fig:2D_maps}). We fit the remaining RVs as in Sect.~\ref{sec:fit_RM} and found no significant changes in the properties derived in Sect.\ref{sec:results_RM}. The contamination from the planet likely does not bias the local stellar RVs beyond the precision of the HARPS data, and the contaminated phase range likely has less influence on the RV model than if it were closer to ingress or egress. Future studies of WASP-121b and similar planets using higher-precision spectrographs should nonetheless take special care with planet-contaminated exposures. The final extraction of the planetary signal was performed using the local RV model derived in Sect.\ref{sec:results_RM} and the new uncontaminated master stellar CCF$_\mathrm{loc}$. Fig.~\ref{fig:2D_atmo_maps} shows the final RV-phase map of the CCF$_\mathrm{atm}$.

\subsection{Results}

\begin{figure}
\centering
\includegraphics[trim=0.5cm 0cm 0.5cm 0.5cm,clip=true,width=\columnwidth]{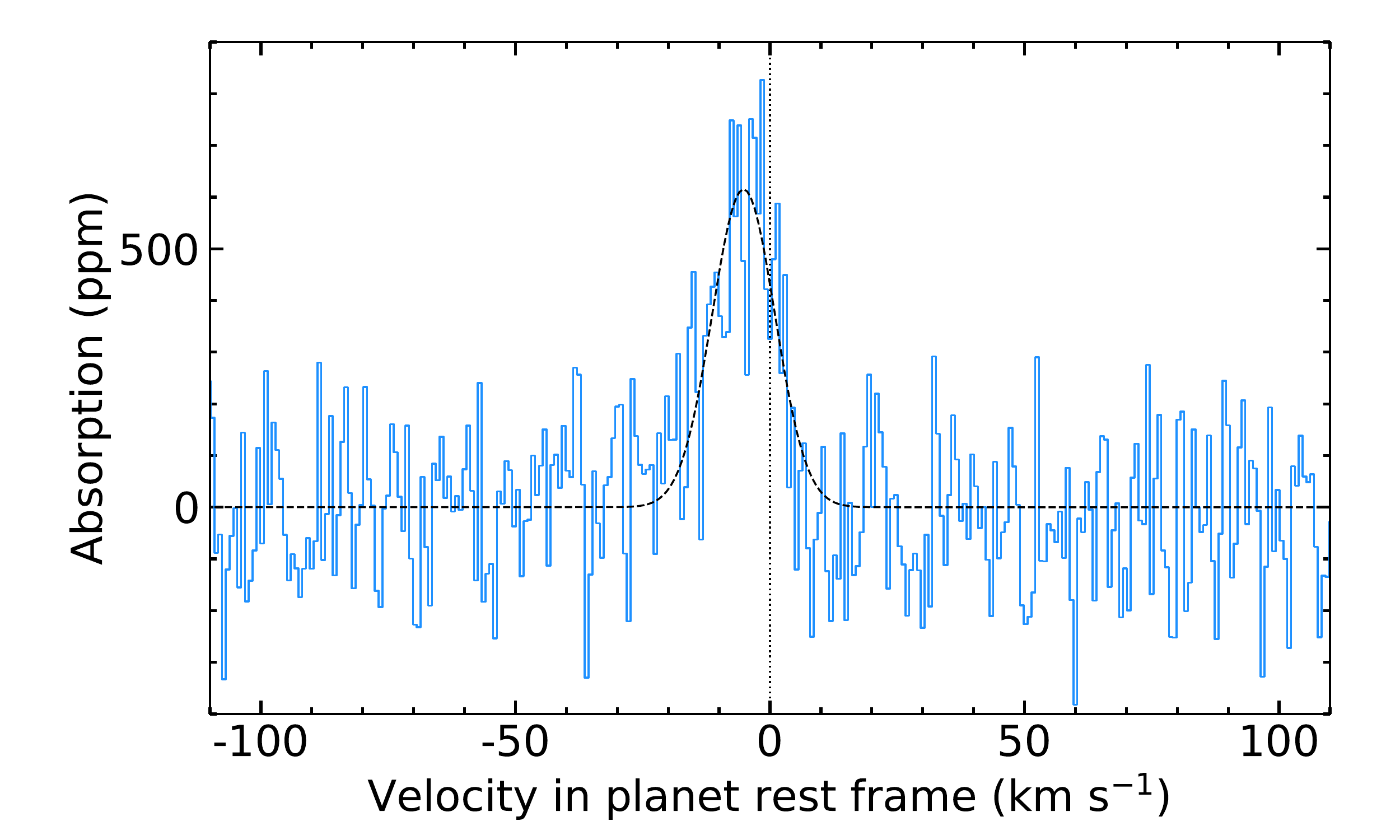}
\caption[]{Master atmospheric CCF$_\mathrm{atm}$ averaged over the full in-transit exposures. The absorption signal from the planetary atmosphere is clearly detected and well approximated by a Gaussian profile (dashed black profile) with a significant blueshift with respect to the planetary rest velocity (vertical dotted black line). }
\label{fig:master_atm}
\end{figure}

The master atmospheric signal, shown in Fig.~\ref{fig:master_atm}, is well fitt with a Gaussian profile. Errors on the master CCF$_\mathrm{atm}$ were set to the dispersion in its continuum. We measure a significant blueshift of -5.2$\pm$0.5\,km\,s$^{-1}$ in the planetary rest frame, and an FWHM of 14.5$\pm$1.2\,km\,s$^{-1}$. Correcting this width for the HARPS LSF broadening (2.61\,km\,s$^{-1}$) and for the blurr induced by the planet motion during an exposure\footnote{The blur does not quadratically broaden the Gaussian profile of the atmospheric signal. We therefore shifted Gaussian profiles at the rate of the planetary motion during a 690\,s long exposure and compared their average with the measured signal.} yields a net FWHM of 12.9$\pm$1.2\,km\,s$^{-1}$ for the atmospheric signal. Thermal broadening contributes negligibly to the measured width (FWHM$_\mathrm{thermal}\sim$1.5\,km\,s$^{-1}$ at 2800\,K). If WASP-121b is tidally locked, then its atmosphere rotates in the stellar rest frame with the same angular velocity as the planet orbits the star (5.7$\times$10$^{-5}$\,rad\,s$^{-1}$). Accounting for the orbital inclination (but assuming that the planet is not inclined with respect to its orbital plane), we obtain a projected rotational RV of 7.15\,km\,s$^{-1}$ for atmospheric layers close to the planet surface, corresponding to an FWHM of 11.9\,km\,s$^{-1}$. Planetary rotation (in the stellar rest frame) therefore likely accounts for most of the atmospheric broadening, especially if the measured signal arises from higher altitude where the planetary rotation induces a higher velocity.\\

The measured blueshift could trace fast winds going from the dayside to the nightside along both terminators, as predicted for atmospheric circulation in the lower atmospheric layers of hot Jupiters (\citealt{Showman2013}). In this scenario the hotspot is expected to be located at the substellar point, as is indeed measured in the TESS phase curve of WASP-121b (\citealt{Bourrier2019}). However, it might then be expected that heat is efficiently restributed through the fast day- to nightside winds, whereas the phase curve revealed a strong temperature contrast. This might indicate that the iron signal arises from different layers than those probed by the TESS photometry. It has been proposed (\citealt{Beatty2019,Keating2019}) that the nightsides of most hot Jupiters are covered with clouds of similar composition, which would form at temperatures of about 1100\,K. With an irradiation temperature of $\sim$3310\,K, WASP-121\,b is in a regime where such clouds are not yet predicted to disperse (\citealt{Keating2019}). The HARPS measurements might therefore probe absorption signals from layers at lower altitudes than are probed by TESS, where fast day- to nightside winds homogenize temperature longitudinally. Meanwhile, the TESS phase curve could trace emission from high-altitude clouds on the nightside (T$_\mathrm{night} <$ 2200\,K at 1$\sigma$), which would hide the emission from the deeper, hotter regions probed on the dayside (T$_\mathrm{day}$ = 2870\,K). Alternatively, the measured blueshift could trace an anisotropic expansion of the upper atmospheric layers, for example, due to the asymmetrical irradiation of the dayside atmosphere (\citealt{Guo2013}) or its compression by stellar wind and radiation. Interestingly, a stronger but marginal blueshift was measured in the metal species escaping WASP-121b (\citealt{Sing2019}), supporting the idea that the atmospheric layers are increasingly blueshifted as their altitude increases. We note that varying the stellar mass within its 3$\sigma$ uncertainties, thus affecting the planet orbital velocity track (e.g., \citealt{Hoeijmakers2019}), does not change the measured blueshift within its uncertainty.

We do not have the precision required to study individual HARPS exposures, but we analyzed the shape and position of the planetary signal averaged over the first half, and then the second half, of the transit (ingress and egress excluded). We found that the absorption signal maintains the same FWHM (13.2$\pm$1.1\,km\,s$^{-1}$ and 13.2$\pm$2.0\,km\,s$^{-1}$, respectively) but becomes more blueshifted (from -3.82$\pm$0.48 to -6.63$\pm$0.86\,km\,s$^{-1}$). Interestingly, blueshifted absorption signals whose shift increases during transit have been observed in the near-IR helium lines of extended planetary atmospheres (\citealt{Allart2018}, \citealt{Nortmann2018}, \citealt{Salz2018}). It is unclear whether these features trace material that escapes from WASP-121b and is blown away by the stellar wind or radiation pressure, as no absorption is observed before or after the transits and the absorption profile shows no strong asymmetries. The atmospheric circulation may show strong spatial asymmetries, and the atmospheric limb probes regions with different speeds as the tidally locked planet rotates during transit. Three-dimensional simulations of the planetary atmosphere and more precise observations are required to explore the origin of the measured blueshift.\\

\section{Conclusion}
\label{sec:conclu}

The ultra-hot Jupiter WASP-121b, transiting a bright F-type star on a near-polar orbit, offers great opportunities to investigate the dynamical and atmospheric properties of giant planets in extreme gravitational and energetic conditions. \\

We combined RV measurements with EulerCam and published TESS photometry to revise the orbital and bulk properties of the planet. Three HARPS transit observations of WASP-121b were then used to refine the orbital architecture of the system. We applied the reloaded RM method to isolate the properties of the stellar photosphere along the transit chord, using a custom mask to compute the CCF of WASP-121, and simultaneous EulerCam photometry to rescale them to their absolute flux level. Analysis of the local RVs from the planet-occulted regions confirms the near-polar orbit of WASP-121b, which leads to a strong degeneracy between impact parameter and stellar rotational velocity. We thus improved the reloaded RM model to include the orbital inclination and semi-major axis in the fit to the local RVs. We further derived independent constraints on the stellar rotation period by analyzing the activity indexes of the star, and by comparing the shapes of the local and disk-integrated stellar lines. This allowed us to derive the stellar inclination, orbital inclination, and 3D obliquity to a high precision (Table~\ref{tab:sys_prop}), and to exclude high differental rotation rates for WASP-121. These measurements will be helpful in constraining studies of WASP-121b past and future dynamical evolution. We encourage follow-up transit observations of the planet to monitor a possible evolution of the obliquity and impact parameter that would result from the nodal precession of the orbit.\\

The custom mask used to calculate the CCFs of WASP-121 was built from the stellar lines, most of which arise from iron transitions. The presence of iron is also expected in the atmosphere of ultra-hot Jupiters because the high temperatures prevent it from condensing. As a result, we developed a new method for removing the contribution of the stellar lines from the local CCFs of the planet-occulted regions and isolating the contribution from the planetary atmosphere. This method is based on the possibility of directly deriving from the data the local stellar lines, uncontaminated by the planet, which is possible when the orbital trajectory of the planet and its transit chord across the stellar surface are sufficiently separated in RV-phase space. The application of this method to the HARPS observations of WASP-121b binned over three transits revealed the absorption CCF of iron in the planet atmospheric limb. The width of the signal is consistent with the rotation of WASP-121b, if it is tidally locked. The absorption signal is blueshifted in the planetary rest frame, increasing from -3.82$\pm$0.48 during the first half of the transit to -6.63$\pm$0.86\,km\,s$^{-1}$ in the second half. This is reminiscent of the effect seen for the ultra-hot gas giant WASP-76\,b (Ehrenreich et al. 2020). These features could arise from day- to nightside winds along both terminators or from the upward winds of an anisotropically expanding atmosphere, combined with the different regions probed by the atmospheric limb as the planet rotates during transit. Observations at higher spectral resolution and with a better sensitivity, for instance, with the ESPRESSO spectrograph, will enable refining the shape of the signal and its temporal evolution. Similar measurements at other wavelengths, searching for species located in different layers than iron, would furthermore allow us to map the full dynamical structure of the WASP-121b atmosphere.\\

Like their colder relatives, ultra-hot Jupiters display a wide range of orbital architectures (from aligned, such as WASP-19b, \citealt{TregloanReed2013} to nearly polar, such as WASP-121b). Ground-based instruments with high resolving power (e.g., HARPS and ESPRESSO in the visible; CARMENES, SPIRou, and NIRPS in the infrared), will make it possible to investigate in details their dynamical properties and to carry out transmission and emission spectroscopy of their atmosphere, allowing us to identify precisely the signatures of their atomic and molecular components and characterize their 3D atmospheric flows.\\


\begin{acknowledgements}
We thank the referee for their fair and useful review of our study. We thank J.B. Delisle for his advice in correcting for activity in the RV measurements and N. Hara for his help in statistical matters. V.B. and R.A acknowledge support by the Swiss National Science Foundation (SNSF) in the frame of the National Centre for Competence in Research ``PlanetS''. This project has received funding from the European Research Council (ERC) under the European Union’s Horizon 2020 research and innovation programme (project Four Aces, grant agreement No 724427; project Exo-Atmos, grant agreement no. 679633). This publication made use of the Data \& Analysis Center for Exoplanets (DACE), which is a facility based at the University of Geneva (CH) dedicated to extrasolar planets data visualisation, exchange and analysis. DACE is a platform of the PlanetS NCCR, federating the Swiss expertise in Exoplanet research. The DACE platform is available at https://dace.unige.ch. N.A-D. acknowledges the support of FONDECYT project 3180063. This work has made use of the VALD database, operated at Uppsala University, the Institute of Astronomy RAS in Moscow, and the University of Vienna.
\end{acknowledgements}

\bibliographystyle{aa} 
\bibliography{biblio} 

\begin{appendix}

\section{Fit to RV data}
\label{apn:RV_fit}

\begin{center}
\begin{figure*}[b!]
\centering
\caption[]{Parameters probed by the MCMC we used to fit the RV measurements of WASP-121 shown in Fig.~\ref{fig:RV_fit}. The maximum likelihood solution, median, mode, and standard deviation of the posterior distribution for each parameter are shown, as well as the 68.27\%, 95.45\%, and 99.73\% confidence intervals. The prior for each parameter can be of type U for uniform, N for normal, or TN for truncated normal. Reference epoch: 2455500.0 BJD.}
\includegraphics[trim=0cm 0cm 0cm 0cm, angle=90,clip=True, scale=0.9]{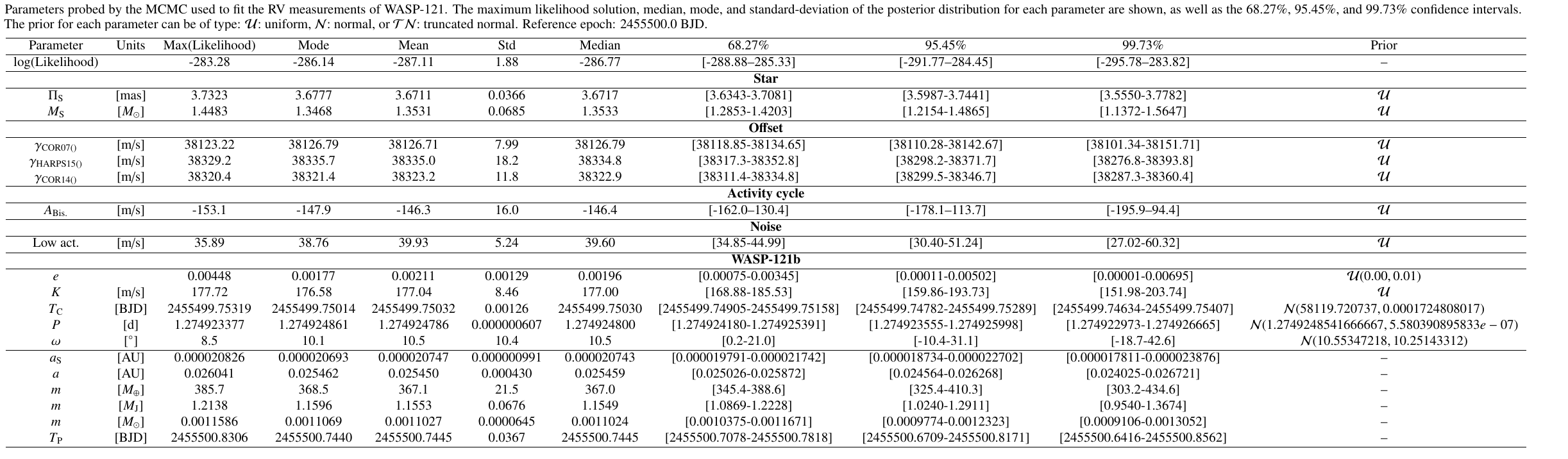}
\label{fig:RV_ana_appendix}
\end{figure*}
\end{center}

\begin{center}
\begin{figure*}[b!]
\centering
\caption[]{Original RV measurements of WASP-121 without any binning}
\includegraphics[trim=0cm 0cm 0cm 0cm, angle=0,clip=True,scale=0.3]{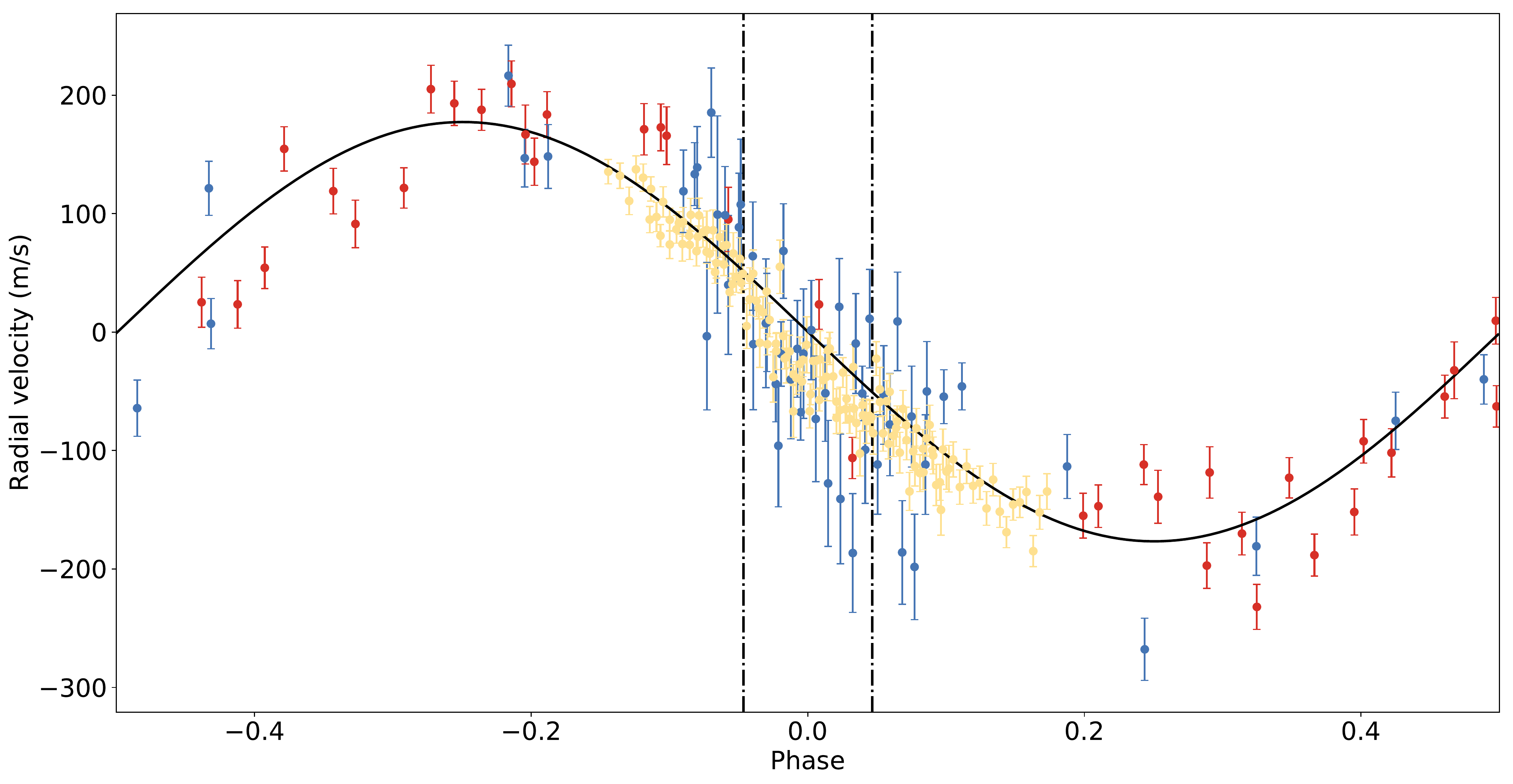}
\label{fig:RV_ana_appendix_nobin}
\end{figure*}
\end{center}

\section{Reloaded RM analysis}
\label{apn:priors_TESS}

\subsection{Effect of a near-polar orbit on determining $v_{\rm eq}\sin i_{*}$ }
\label{apn:polar_orb}
 
During the transit, when the orbital phase $\phi\sim$0, the local RVs of the stellar surface regions that are occulted by the planet approximate as (see Eq. 4 in \citealt{bourrier2015_koi12} or Eq. 8 in \citealt{Cegla2016})
\begin{equation}
v_{\rm loc}(\phi)= v_{\rm eq}\sin i_{*} \, \left( \frac{a_{p}}{R_{\star}} \, \cos(\lambda)\,2\,\pi\,\phi\,+\,b\,\sin(\lambda) \right). 
\end{equation}
For misaligned orbits when $\lambda$ is close to $\pm$90$^{\circ}$, the local RVs further approximate as
\begin{equation}
v_{\rm loc}(\phi)= v_{\rm eq}\sin i_{*} \, \frac{a_{p}}{R_{\star}} \, (\frac{\pi}{2} - \lambda) \,2\,\pi\,\phi\,+\,b\,v_{\rm eq}\sin i_{*}. 
\end{equation}
The level of the local RV series is therefore controlled at first order by $v_{\rm loc}(0)= b\,v_{\rm eq}\sin i_{*}$. For a planet on a near-polar orbit and with a low impact parameter such as WASP-121b, small variations in $b$ thus have a direct and strong effect on $v_{\rm eq}\sin i_{*}$. Conversely, the slope of the local RVs variation with phase can be expressed as
\begin{equation}
\dot{v}_{\rm loc}= 2\,\pi\,v_{\rm loc}(0)\,\frac{(\frac{\pi}{2} - \lambda)}{\cos(i_{\rm p})} = 2\,\pi\,v_{\rm loc}(0) \, \frac{(\frac{\pi}{2} - \lambda)}{(\frac{\pi}{2} - i_{\rm p})}. 
\end{equation}
Here $i_{\rm p}$ is close to 90$^{\circ}$. Therefore, a variation in the orbital inclination directly translates into a variation of the sky-projected obliquity, with a factor $\dot{v}_{\rm loc}/2\,\pi\,v_{\rm loc}(0)$ that can be low for near-polar orbits. This is the case for WASP-121b, with $v_{\rm loc}(0)\sim$1.4\,km\,s$^{-1}$ and $\dot{v}_{\rm loc}\sim$18\,km\,s$^{-1}$ per phase unit, yielding a factor of about 2. A near-polar orbit is therefore an adverse case for the interpretation of the local RVs in terms of stellar rotational velocity, but may not prevent a precise derivation of the sky-projected obliquity.  \\

\subsection{Posterior probability distributions}
\label{apn:pdf_RM}

\begin{center}
\begin{figure*}[b!]
\centering
\includegraphics[trim=0cm 0cm 0cm 0cm, clip, scale=0.95]{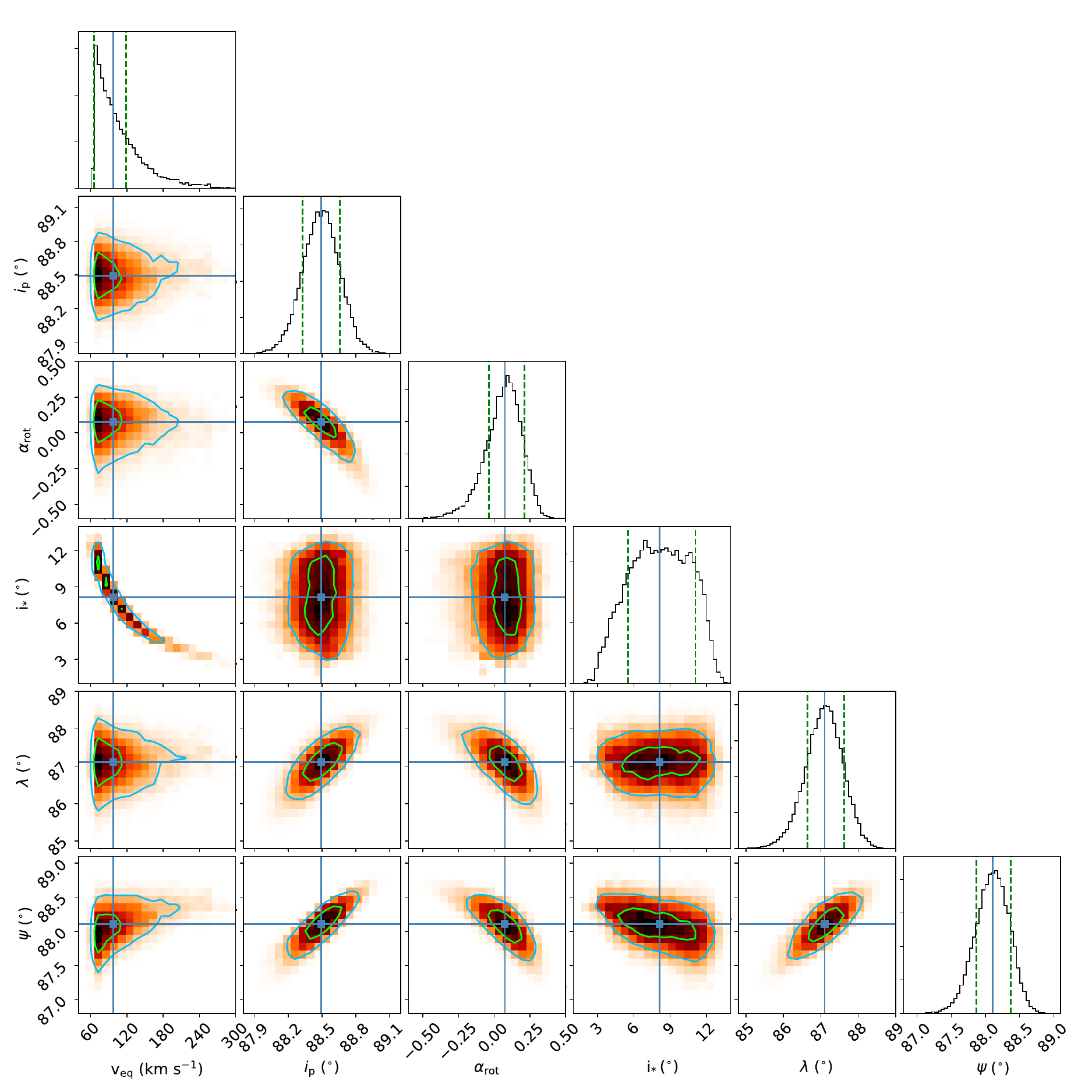}
\caption[]{Correlation diagrams for the probability distributions of the updated reloaded RM model parameters. Samples have been limited to the region with low stellar inclination. The inner green and outer light blue contours show the 1 and 2$\sigma$ simultaneous 2D confidence regions that contain 39.3\% and 86.5\% of the samples, respectively. 1D histograms correspond to the distributions projected on the space of each line parameter. The deep blue lines indicate their median values, with dashed green lines showing the 1$\sigma$ highest density intervals. $\psi$ is derived from other parameters and is not an MCMC jump parameter.}
\label{fig:PD_lowi}
\end{figure*}
\end{center}

\begin{center}
\begin{figure*}[b!]
\centering
\includegraphics[trim=0cm 0cm 0cm 0cm, clip, scale=0.95]{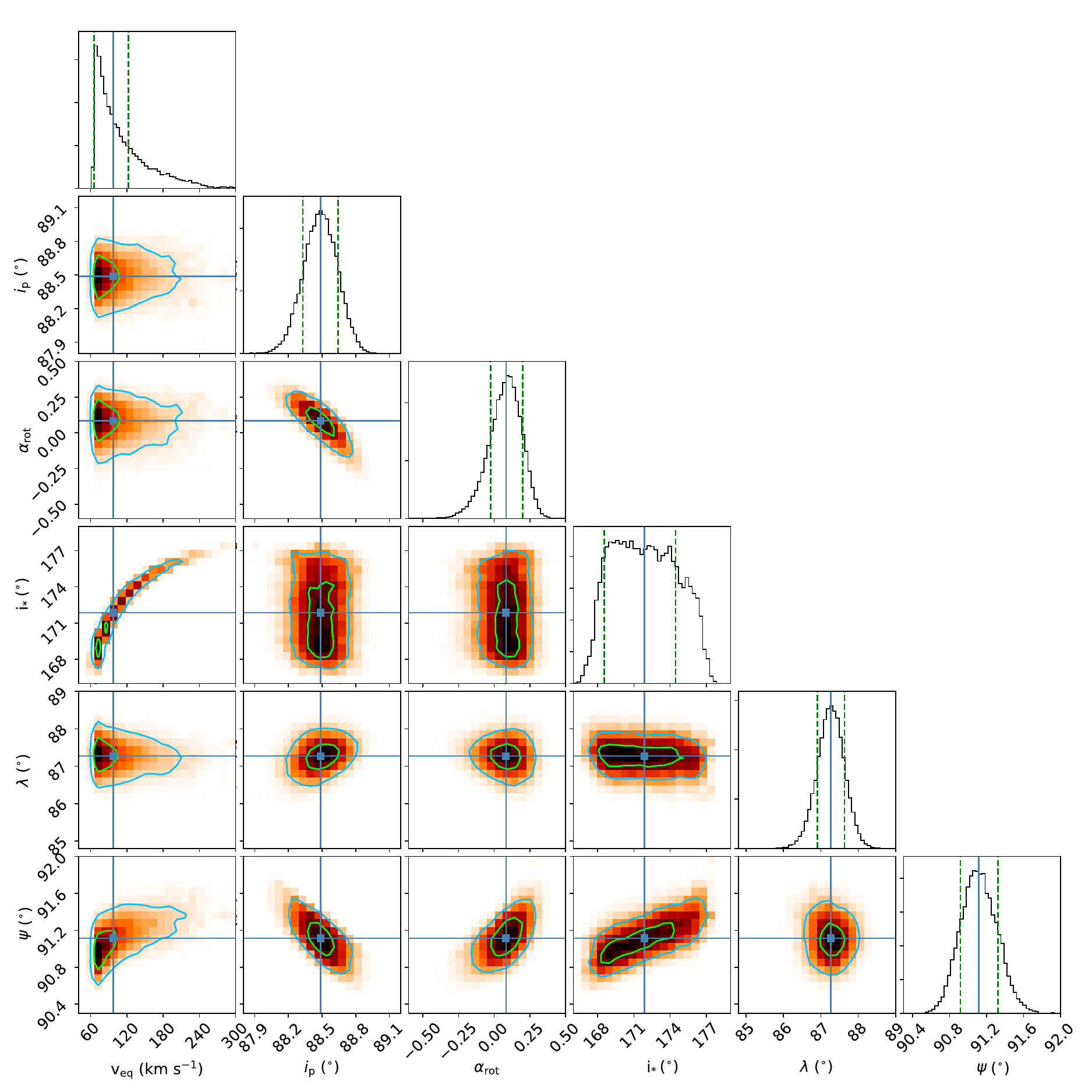}
\caption[]{Same as in Fig.~\ref{fig:PD_lowi}, with samples limited to the region with high stellar inclination.} 
\label{fig:PD_highi}
\end{figure*}
\end{center}

\end{appendix}

\end{document}